\documentclass[preprint,showpacs,longbibliography]{revtex4-1}
\usepackage{amsmath, amssymb, appendix, enumerate, hyperref, graphicx, subfigure}
\usepackage{pdfpages}
\usepackage{color}

\let\origleft\left
\let\origright\right

\renewcommand{\left}{\mathopen{}\mathclose\bgroup\origleft}
\renewcommand{\right}{\aftergroup\egroup\origright}

\begin{document}
\title{Nonadiabatic Coulomb effects in strong-field ionization in circularly polarized laser fields}
\author{Jivesh Kaushal}
\author{Olga Smirnova}
\affiliation{Max Born Institute, Max Born Strasse 2a, 12489 Berlin, Germany}
\begin{abstract}
We develop the recently proposed analytical $R$-matrix (A$R$M) method to encompass strong field ionization by circularly polarized fields, for atoms with arbitrary binding potentials. Through the A$R$M method, the effect of the core potential can now be included consistently both during and after ionization. We find that Coulomb effects modify the ionization dynamics in several ways, including modification of (i) the ionization  times, (ii) the initial conditions for the electron continuum dynamics, (iii)  the ``tunneling angle," at which the electron ``enters" the barrier,  and (iv) the electron drift momentum. We derive analytical expressions for the Coulomb-corrected ionization times, initial velocities,  momentum shifts, and  ionization rates in circularly polarized fields, for arbitrary angular momentum of the initial state. We also analyze how nonadiabatic Coulomb effects  modify (i) the calibration of the attoclock in the angular streaking method and  (ii)  the ratio of ionization rates from $p^{-}$ and $p^{+}$ orbitals, predicted by I. Barth and O. Smirnova [Phys. Rev. A \textbf{84}, 063415 (2011)] for short-range potentials.

\end{abstract}
\maketitle

\section{Introduction}

Single and double ionization in circularly polarized strong laser fields is a sensitive probe of attosecond dynamics \cite{eckle2008,eckle2008-2,pfeiffer2011,pfeiffer2012,akagi2009,fleischer2011}. Strong-field ionization is often viewed as electron tunneling from atoms and molecules through the barrier created by the laser field and the core potential. The adiabatic approximation, frequently used to describe tunneling, implies quasistatic electric field and zero electron velocity immediately after ionization (at the tunnel exit). This adiabatic picture is used for the interpretation of current experiments in circularly polarized laser fields \cite{eckle2008, eckle2008-2, pfeiffer2011, pfeiffer2012, akagi2009, fleischer2011} within the two-step model. This model merges quantum and classical approaches by combining (i) the adiabatic approximation for the quantum ionization step with (ii) the classical trajectories calculation after tunneling. In this second step, an ensemble of classical trajectories is launched outside the barrier; the distribution of initial velocities parallel and perpendicular to the direction of the instantaneous laser field is centered around 0, as predicted by the adiabatic tunneling theory.

However, strictly speaking, in circularly polarized laser fields the tunneling barrier is rotating. This rotation manifests itself in the nonadiabatic electron response, which becomes significant in the regime of the Keldysh parameter $\gamma \geq 1$. Nonadiabatic effects change tunelling from essentially one-dimensional, characteristic of the static limit \cite{landau-lifshitz, pfeiffer2012}, to two-dimensional. As shown in \cite{barth2011}, for short-range potentials substantial deviations from the adiabatic approximation arise already for $\gamma^2 \simeq 0.5$, which also questions the validity of this approximation for the long-range core potentials under similar conditions. We note that $\gamma^2 \simeq 0.5$ is a typical regime for recent experiments with laser radiation around 1600$-$1300 nm and systems with ionization potential $I_p \sim 10$ eV (see, e.g., \cite{torres2010}).

Here we provide a rigorous analytical framework for treating the effects of long-range potential and laser field on equal footing and include nonadiabatic effects due to the long range potential. In particular, we show that nonadiabatic Coulomb effects lead to a non-zero initial velocity both parallel and perpendicular to the direction of the instantaneous laser field even for the central electron trajectory (for which the ionization rate maximizes), when it emerges from the classically forbidden region.

We find that the nonadiabatic Coulomb effects modify the ionization dynamics in several ways, including modification of (i) the ionization (exit) times, (ii) the initial conditions for electron continuum dynamics, (iii) the ``tunneling angle," at which the electron ``enters" the barrier, and (iv) the electron drift momentum. We derive analytical expressions for the ionization times, initial velocities, momentum shifts, and ionization amplitudes and rates in circularly polarized fields for arbitrary angular momentum of the initial state. We also analyze how the nonadiabatic Coulomb effects modify (i) the calibration of the attoclock in the angular streaking method \cite{eckle2008, eckle2008-2, pfeiffer2011, pfeiffer2012}, and (ii) the ratio of ionization rates from $p^{-}$ and $p^{+}$ orbitals obtained for short-range potentials in \cite{barth2011}.

Our tool is the gauge-invariant, analytical $R$-matrix (A$R$M) method, which we have recently developed \cite{lisa2012} for linearly polarized fields. The strength of the A$R$M method is the ability to treat consistently the effects of long-range potential and the laser field \cite{lisa2012} as well as multielectron effects \cite{lisa2012-2}. The main idea of the $R$-matrix method was adopted from the study of collision processes and nuclear resonance reactions \cite{r-matrix}, where the primary purpose was to isolate the strongly interacting kernel from the region where these interactions were significantly weaker and can be considered in asymptotic approximation. In this sense, the $R$-matrix approach developed in collision physics meets a crucial requirement of strong-field physics, in that it can be used to separate the region of the configuration space, where the Coulomb forces are much stronger than the laser field, from the outer region, where the Coulomb potential quickly becomes almost negligible compared to the driving laser field. We note that fully numerical time-dependent $R$-matrix approach for strong-field dynamics was developed and successfully applied in \cite{lyasght2011}.

In \cite{lisa2012}, a detailed analysis and benchmarking of the A$R$M method was provided for strong linearly polarized laser fields, including the derivation of analytical results for the instantaneous ionization amplitudes and the sub-cycle ionization rates for single-active-electron systems. With suitable approximations, the results for the cycle-averaged ionization rates from \cite{lisa2012} agree with those obtained by Perelomov, Popov, and Terent\'ev (PPT) \cite{ppt1966}. For hydrogen-like atoms and ions, the PPT rates (with the new correction factor derived in \cite{popruzhenko2008}) were shown to be accurate for arbitrary values of the Keldysh parameter.

The difference between our problem in circularly polarized fields and the case of the linearly polarized fields analyzed in \cite{lisa2012} lies in the fundamentally two-dimensional character of tunneling, i.e., the time-varying angles between the position vectors, the electron velocity $\mathbf{v}_{\mathbf{p}}(t)$, and the laser vector potential $\mathbf{A}(t)$. As a consequence, certain approximations in \cite{lisa2012} that were helpful in deriving a physically transparent solution for the linearly polarized field are not always adequate in the case of circularly polarized fields. Here, we refine the A$R$M method and extend it to strong circularly polarized fields. We show that an appropriate choice of the boundary between the inner and the outer regions allows one to build the hierarchy of interactions and show how the long-range effects can be included consistently within the iterative approach.

Our strategy can be summarized as follows. Following \cite{lisa2012}, we introduce an ``$R$-matrix" sphere of radius $a$, which splits the configuration space into the inner and outer regions.
\begin{description}
	\item[{\it A. The inner region}] In the inner region the Coulomb field dominates, and the effects of the laser field on the inner region wave function can be included in the quasistatic approximation. Further approximation, such as using the field-free wave function is justified for fields significantly smaller than $\kappa^3$, where $\kappa = \sqrt{2I_p}$, $I_p$ is the ionization potential.
	\item[{\it B. The boundary}] The Bloch operator is used to ``pass"  the information about the electron wave function from the inner region to the outer region. The outer region Green's function is used to propagate the outer region wave function from this boundary to the detector. The boundary value is given by the inner region wave function at the surface of the ``$R$-matrix" sphere. Boundary matching ensures that the final result does not depend on the choice of the boundary value $a$.
	\item[{\it C. The outer region}] In the outer region, the Coulomb potential is weak and can be included in the Eikonal-Volkov approximation (EVA) \cite{olga2008}. It has been shown previously \cite{olga2006}, that the EVA is adequate for the soft-core potentials, i.e., for long-range potentials outside the singularity region.
	\item[{\it D. Propagation in the outer region}] This involves integration over the surface of the $R$-matrix sphere ($\theta'$, $\phi'$) and over all times ($t'$) of ``transition" through the boundary. Due to the large action $S$ of the electron in the strong laser field, the integrals are taken from the highly oscillating function $P(\theta',\phi',t')e^{-iS(\theta',\phi',t')}$ and are accumulated in the vicinity of their respective stationary (saddle) points $\theta_s'$, $\phi_s'$, and $t_s'$, defined by the solutions of the equations $\partial_{\theta'}S = 0$, $\partial_{\phi'}S = 0$, and $\partial_{t'}S = 0$, respectively, where the subscript denotes the derivative with regard to (w.r.t.) that variable. The action is given by $S = S^{\text{SFA}} + G_C$, where the strong-field-approximation (SFA) action is associated with the dynamics in the laser field and short-range potential, and $G_C$ is the action associated with the interaction with the long-range potential of the core under the EVA \cite{olga2008} and describes Coulomb-laser coupling \cite{olga2007}. Since only a vicinity of saddle points contributes to the integral, we do a Taylor expansion of $G_C$ around SFA saddle points $\theta_s^{\prime(0)}$, $\phi_s^{\prime(0)}$, and $t_s^{\prime(0)}$. After this expansion the integral over the surface of the sphere is calculated exactly. The integral over $t'$ is evaluated using the saddle-point method. The actual, full saddle point $t_s'$ (shifted from $t_s^{\prime(0)}$ due to long-range effects) is found within the iterative approach. Formally, nonadiabatic effects in ionization rates arise due to the deviations from the stationary trajectory included via Taylor expansion of $G_C$. Nonadiabatic Coulomb effects also manifest itself in the photoelectron spectra and will be considered in our subsequent paper \cite{lisa2013}.
	\item[{\it E. Iterative approach to saddle-point equation for ${\mathbf{t'}}$}] By construction, in the outer region $G_C$ presents a perturbation to the SFA action $S^{\text{SFA}}$ and therefore can only slightly shift the SFA saddle point $t_{s}^{\prime(0)}$, which corresponds to the stationary SFA action: $\partial_{t'}S^{\text{SFA}} = 0$. Thus, as a first correction to the saddle point, due to the interaction with the long-range potential, $t_s^{\prime(1)} = t_{s}^{\prime(0)} + \Delta t_s^{\prime(0)}$, where $\Delta t_s^{\prime(0)}$ can be found by iterations with respect to $G_C$. The first iteration includes only linear terms in $\Delta t_{s}^{\prime(0)} \sim \mathcal{O}(G_C)$. In our approach we keep only the first-order correction terms  consistently throughout. The saddle-point equation $\partial_{t'}S = 0$ can be expanded around $t_{s}^{\prime(0)}$: $\partial_{t'}S^{\text{SFA}}\left(t_{s}^{\prime(0)}\right) + \Delta t_s^{\prime(0)}\partial_{t'}^2S^{\text{SFA}}\left(t_{s}^{\prime(0)}\right) + \partial_{t'}G_C\left(t_{s}^{\prime(0)}\right) = 0$, yielding $\Delta t_{s}^{\prime(0)} = -\frac{\partial_{t'}G_C\left(t_s^{\prime(0)}\right)}{\partial_{t'}^2S^{\text{SFA}}\left(t_s^{\prime(0)}\right)}$. Note that since the SFA action is stationary, $\partial_{t'}S^{\text{SFA}}\left(t_{s}^{\prime(0)}\right) = 0$, the shift due to $\Delta t_s^{\prime(0)}$ will only change the value of the SFA action in the second order w.r.t. $G_C$: $S\left(t_{s}^{\prime(1)}\right)=S^{\text{SFA}}\left(t_{s}^{\prime(0)}\right)+
G_C\left(\theta_{s}^{\prime(0)},\phi_{s}^{\prime(0)},t_{s}^{\prime(0)}\right) + \mathcal{O}\left(\left(\Delta t_{s}^{\prime(0)}\right)^2\right)$.
However, $t_{s}^{\prime(1)}$ will contribute to the pre-exponential factor $P(\theta',\phi',t')$ in the integral.
	\end{description}

Below we detail our method and show how it can be used to obtain ionization amplitudes and ionization rates using both the time-domain and the frequency-domain approaches. The time-domain approach is technically simpler and allows one to consider temporal dynamics of ionization, including the time evolution of electron momentum distributions \cite{lisa2013} and ionization rates.

The paper is organised as follows. Section~\ref{section:wave function} introduces basic equations. Section~\ref{section:time_domain} develops the time-domain approach. Section~\ref{section:phys_pic} discusses the physical picture arising from the theory developed in Secs.~\ref{section:wave function}$-$\ref{section:time_domain}. In Sec.~\ref{section:phys_pic}, we describe modifications of the ionization dynamics due to Coulomb effects. These include (i) Coulomb corrections to ionization times, (ii) initial conditions for electron continuum dynamics, (iii) calibration of the attoclock in the angular streaking method, and (iv) Coulomb corrections to the ``tunneling angle," including the Coulomb corrections to the ratio of ionization rates from $p^{-}$ and $p^{+}$ orbitals obtained for the short range potentials in \cite{barth2011}. Section~\ref{section:conclusion} concludes the work. Appendix~\ref{app:boundary} presents additional calculations related to the boundary matching. Appendix~\ref{app:freq_domain} develops the frequency-domain approach, pioneered in the PPT work on short-range potentials. This approach requires more involved algebra but allows the most straightforward connection to the PPT results. Appendix~\ref{app:sub_time_domain} extends the time-domain method in Sec.~\ref{section:time_domain} to introduce observables characterizing subcycle ionization dynamics. Appendixes~\ref{app:limit_bessel} and \ref{app:equivalence} present miscellaneous calculations.

\section{Basic equations} \label{section:wave function}

Following \cite{lisa2012}, we introduce the Bloch operator $\hat{L}^{\pm}\left(a\right)$ to split the configuration space into the inner and outer regions. Parameter $a$ represents the radius of the $R$-matrix sphere; the inner region is inside the sphere, the outer region is outside of the sphere. The standard Hamiltonian $\hat{H}$ including both Coulomb $ V_C(\mathbf{r})$ and laser-field interaction $V_L(t)$,
	\begin{equation}
		\hat{H} = \frac{\hat{\mathbf{p}}^2}{2} + V_C(\mathbf{r}) + V_L(t),
	\end{equation}
used in the Schordinger equation,
	\begin{equation}
		i\frac{\partial\psi\left(\mathbf{r},t\right)}{\partial t} = \hat{H}\psi\left(\mathbf{r},t\right).
	\end{equation}
	\begin{equation}
		\psi\left(\mathbf{r},t=t_0\right) = \psi_g\left(\mathbf{r}\right),
	\end{equation}
can be modified to
	\begin{equation}
			i\frac{\partial\psi\left(\mathbf{r},t\right)}{\partial t} = \hat{H}_B^{(\pm)}\psi\left(\mathbf{r},t\right) - \hat{L}^{(\pm)}(a)\psi\left(\mathbf{r},t\right), \label{wave_equation}
	\end{equation}
where $\hat{H}_B^{(\pm)}=\hat{H} + \hat{L}^{(\pm)}(a)$. Following arguments developed in \cite{lisa2012}, we can express the solution in the outer region via the solution in the inner region as
	\begin{equation}
		|\psi_{\text{out}}(t)\rangle = i\int_{t_0}^tdt'\,\hat{U}_B^{(-)}(t,t')\hat{L}^{(-)}(a)|\psi_{\text{in}}(t')\rangle, \label{non-homogeneous}
	\end{equation}
where for the outgoing solution, we use $\hat{L}^{(-)}(a)$ and the governing equation for the evolution operator $\hat{U}_B^{(-)}(t,t')$ is
	\begin{equation}
		i\frac{\partial}{\partial t}\hat{U}_B^{(-)}(t,t')= \hat{H}_B^{(-)}(t)\hat{U}_B^{(-)}(t,t').
	\end{equation}
In our time-domain approach, detailed in the next section, we start the analysis from the expression for the ionization amplitude $a_{\mathbf{p}}(T)=\langle \mathbf{p}|\psi_{\text{out}}(T)\rangle$ (see \cite{lisa2012} for discussion)
	\begin{equation}
		a_{\mathbf{p}}(T) = i\int_{t_0}^Tdt'\int_{}^{}d\mathbf{r}'\int_{}^{}d\mathbf{r}''\,\langle\mathbf{p}\vert\hat{U}^{(-)}_B(T,t')\vert\mathbf{r}'\rangle\langle\mathbf{r'}
\vert\hat{L}^{(-)}(a)\vert\mathbf{r}''\rangle\langle\mathbf{r}''\vert\psi_{\text{in}}(t')\rangle. \label{a_pTdef}
	\end{equation}
Taking into account the explicit form of the Bloch operator in coordinate representation,
	\begin{align}
		\langle\mathbf{r'}\vert\hat{L}^{(-)}(a)\vert\mathbf{r}''\rangle &= \delta(r - a)\delta(\mathbf{r'}-\mathbf{r''})\hat{B},\\
		\hat{B}\psi\left(\mathbf{r},t\right) &= \left.\left(\frac{d}{dr} + \frac{1}{r}\right)\psi\left(\mathbf{r},t\right)\right|_{r=a}, \label{L_coord}
	\end{align}
we can rewrite Eq.~\eqref{a_pTdef} as
	\begin{equation}
		a_{\mathbf{p}}(T) = i\int_{t_0}^Tdt'\int_{}^{}d\mathbf{r}'\,G_{B}^{(-)}(\mathbf{p},T;\mathbf{r'},t')\delta(r'-a)B(a,\theta',\phi',t'), \label{a_pTdef1}
	\end{equation}
where the function $B(a,\theta',\phi',t')$ represents the inner-region wave function at the partition surface $r'=a$ for all times $t_0<t'<T$,
	\begin{equation}
		\begin{split}
			B(a,\theta,\phi,t') & = \left.\left(\frac{d}{dr} + \frac{1}{r}\right)\psi_{\text{in}}(\mathbf{r},t')\right|_{r=a}, \label{B_def}
		\end{split}
	\end{equation}
and $G_{B}^{(-)}(\mathbf{p},T;\mathbf{r}',t') = \left\langle\mathbf{p}\left\vert\hat{U}_B^{(-)}(T,t')\right\vert\mathbf{r}'\right\rangle$ is the Green's function for the modified Hamiltonian $\hat{H}_{B}^{(-)}$ for propagating from the boundary $r' = a$ instead of the origin. As shown in \cite{lisa2012}, the error incurred in approximating this exact Green function with the Eikonal-Volkov approximated Green's function $G^{\text{EVA}}(\mathbf{p},T;\mathbf{r}',t') $ defined on the EVA states \cite{olga2008},
	\begin{equation}
			G^{\text{EVA}}(\mathbf{p},T;\mathbf{r}',t') = \frac{1}{(2\pi)^{3/2}}\,e^{-i\mathbf{v}_{\mathbf{p}}(t')\cdot\mathbf{r}'-\frac{i}{2}\int_{t'}^{T}d\tau\,v_{\mathbf{p}}^2(\tau)}
e^{i\int_{T}^{t'}U(\mathbf{r}_L(\tau;\mathbf{r}',\mathbf{p},t'))}e^{-iG_{0\mathbf{p}}(\mathbf{r}_L(T;\mathbf{r}',\mathbf{p},t'))}, \label{GpTdef}
	\end{equation}
is exponentially small. In the above expression we have defined
	\begin{equation}
		\mathbf{r}_L(\tau;\mathbf{r},\mathbf{p},t) = \mathbf{r} + \int_{t}^{\tau}d\zeta\,\mathbf{v}_{\mathbf{p}}(\zeta), \label{r_L}
	\end{equation}
the characteristic trajectory along which the Coulomb correction is calculated as a perturbation to the Volkov electron \cite{olga2008} and $\mathbf{v}_{\mathbf{p}}(t) = \mathbf{p} + \mathbf{A}(t)$ is the kinetic momentum.

\section{The Time-domain approach} \label{section:time_domain}

We use Eq.~\eqref{a_pTdef1} for the ionization amplitude and Eq.~\eqref{GpTdef} to obtain
	\begin{equation}
		a_{\mathbf{p}}(T) = \frac{i}{(2\pi)^{3/2}}\int_{t_0}^Tdt'\int_{}^{}d\mathbf{r}'\,e^{-i\mathbf{v}_{\mathbf{p}}(t')\cdot\mathbf{r}'}e^{-\frac{i}{2}\int_{t'}^{T}d\tau\,v_{\mathbf{p}}^2(\tau)}e^{-iG_C(\mathbf{p},T; \mathbf{r}', t')}\delta(r'-a)B(a,\theta',\phi',t'), \label{a_pTdef2}
	\end{equation}
with the Coulomb phase term defined as
	\begin{equation}
		G_C(\mathbf{p},T;\mathbf{r}',t')=\int_{t'}^{T}d\tau\,U(\mathbf{r}_L(\tau;\mathbf{r'}, \mathbf{p}, t')). \label{G}
	\end{equation}
Since the time $T$ of observation is sufficiently far, so that we can consider $T \to \infty$ for all practical purposes, we have made the approximation as in \cite{lisa2012}, ignoring the distortions of the phase front from the plane wave, $G_{0\mathbf{p}}\to0$ \cite{olga2008} in Eq.~\eqref{GpTdef}.

\subsection{Transition through the boundary ${r' = a}$} \label{subsection:boundary}

The function $B(a, \theta', \phi', t')$ reflects the value of the inner-region wave function at the boundary $r' = a$. In the inner region the Coulomb field dominates and the effects of the laser field on the inner region wave function $\psi_{\text{in}}(\mathbf{r'},t')$ can be included in the quasistatic approximation. Following \cite{lisa2012}, the boundary is placed in the asymptotic region $E_0a/I_p \ll 1 \ll \kappa a$ of the ground-state wave function, where $\kappa = \sqrt{2I_p}$, and $E_0$ is the amplitude of the laser field. In this region the error in approximating the polarized wave function with the field-free initial wave function is of the order of $\sim E_0 a^2/\kappa$ \cite{murray2010}. Thus, for sufficiently weak fields such that $E_0/\kappa^3 \ll 1/a^2\kappa^2$ the inner region wave function $\psi_{\text{in}}(\mathbf{r'},t')$ can be substituted by the field-free bound-state wave function, without affecting the boundary matching. This approximation was first used in the PPT method \cite{ppt1966}. The asymptotic radial part of this wave function is given by
	\begin{equation}
		\varphi_{\kappa\ell}\left(r'\right) = C_{\kappa\ell}\kappa^{3/2}\frac{e^{-\kappa r'}}{\kappa r'}(\kappa r')^{Q/\kappa}. \label{bound_approx}
	\end{equation}
Due to the invariance of the boundary term under the addition of a function $b_0/r'$, we can choose $b_0$ appropriately to get
	\begin{equation}
		\begin{split}
			B(a,\theta',\phi',t') &= \left.\left(\frac{d}{dr'}-\frac{b_0-1}{r'}\right)\varphi_{\kappa\ell}(r')\right|_{r'=a}\\
			&= -\kappa\varphi_{\kappa\ell}(a) \label{boundary}
		\end{split}
	\end{equation}
for $b_0 = Q/\kappa$. Using Eq.~\eqref{boundary} and evaluating the Delta function over $r'$,
	\begin{equation}
		\begin{split}
			a_{\mathbf{p}}(T) = \frac{i\kappa a^{2}}{(2\pi)^{3/2}}\int_{t_0}^Tdt'\int_{}^{}d \Omega'\,e^{-i\mathbf{v}_{\mathbf{p}}(t')\cdot\mathbf{a}}e^{-iS^{\text{SFA}}(\mathbf{p},T; t')}e^{-iG_C(\mathbf{p},T; \mathbf{a}, t')}\\\varphi_{\kappa\ell}(a) N_{\ell m}P_{\ell}^m(\cos\theta')e^{im\phi'}, \label{a_pTdef3}
		\end{split}
	\end{equation}
where $N_{\ell m} = \sqrt{\frac{2\ell+1}{4\pi}\frac{(\ell-|m|)!}{(\ell+|m|)!}}$, $\mathbf{a} = a(\sin\theta'\cos\phi'\,\hat{\mathbf{x}} + \sin\theta'\sin\phi'\,\hat{\mathbf{y}} + \cos\theta'\,\hat{\mathbf{z}})$. We use $\mathbf{A}(t) = -A_0(\cos\omega t\,\hat{\mathbf{x}} + \sin\omega t\,\hat{\mathbf{y}})$.
The SFA phase is given by
    \begin{equation}
	    S^{\text{SFA}}(\mathbf{p},T;t') = \frac{1}{2}\int_{t'}^{T}d\tau\,v_{\mathbf{p}}^2(\tau) - \frac{\kappa^2}{2}(t'-t_0). \label{S_SFA}
    \end{equation}

Note that in the outer region, the long-range interaction of the electron with the core is described by the phase term $G_C(\mathbf{p},T; \mathbf{r}', t')$, and involves integration of the Coulomb potential along the electron trajectory in the laser field. The trajectory originates from point $\mathbf{a}$ on the boundary at time $t'$ [Eq.~\eqref{r_L}].

\subsection{Iterative approach to solution of saddle point equations}

The calculation of the ionization amplitude $a_{\mathbf{p}}(T)$ involves integration over all starting points of the trajectory on the sphere and all times $t'$ of ``transition" through the boundary $r' = a$. The integrand of Eq.~\eqref{a_pTdef3} can be written in the form $P(\theta', \phi', t')e^{-iS(\theta', \phi', t')}$, where the prefactor $P(\theta', \phi', t')$ reflects the value of the inner-region wave function at the boundary $r' = a$, and the exponent is given by the electron action. The action $S(\theta', \phi', t')$ consists of two parts, $S=S^{\text{SFA}} + G_C$, where $S^{\text{SFA}}$ is the SFA action [Eq.~\eqref{S_SFA}], associated with the ionization dynamics in the short-range potential, and $G_C$ [Eq.~\eqref{G}] is the term responsible for the long-range interaction with the core, describing the Coulomb-laser coupling \cite{olga2007}.

Due to the large value of the action $S$ for the electron in a strong laser field, the integrals are accumulated in the vicinity of stationary (saddle) points $\theta_s'$, $\phi_s'$, and $t_s'$, which satisfy the equations
	\begin{align}
		\frac{\partial S}{\partial\theta'} = 0, \quad \frac{\partial S}{\partial\phi'} = 0, \quad \frac{\partial S}{\partial t'} + a\frac{\partial v_{\mathbf{p}}(t')}{\partial t'} = 0, \label{saddle_point_equations}
	\end{align}
where the saddle point in time has an additional term $av_{\mathbf{p}}(t')$, which, as we show in Appendix~\ref{subapp:wave function}, is a result of propagation from a finite boundary, and comes from the exact evaluation of the surface integral.

To solve these equations, we recall that, by construction, in the outer region $G_C$ presents a perturbation to $S^{\text{SFA}}$ and therefore can only slightly shift the SFA saddle points $\theta_{s}^{\prime(0)}$, $\phi_{s}^{\prime(0)}$, and $t_{a}^{\prime(0)}$ satisfying the equations
	\begin{align}
		\frac{\partial S^{\text{SFA}}}{\partial\theta'} = 0, \quad \frac{\partial S^{\text{SFA}}}{\partial\phi'}=0, \quad \frac{\partial S^{\text{SFA}}}{\partial t'} + a\frac{\partial v_{\mathbf{p}}(t')}{\partial t'} = 0 \label{SFA_saddle_point_equations}
   \end{align}
Thus, the saddle points for the total action $S = S^{\text{SFA}} + G_C$ can be written as
	\begin{align}
		\theta_s' = \theta_{s}^{\prime(0)} + \Delta \theta_s', \quad \phi_s' = \phi_{s}^{\prime(0)} + \Delta \phi_s', \quad t_a' = t_a^{\prime(0)} + \Delta t_a',
   \end{align}
where $\Delta\theta_s'$, $\Delta\phi_s'$, and $\Delta t_a'$ are the small corrections to the SFA saddle points and can be found perturbatively. Subscript ``$a$" in $\Delta t_a'$ indicates that the time $\Delta t_a'$ and the time $t_a^{\prime(0)}$ are affected by the position of the boundary due to the boundary-dependent term $a\frac{\partial v_{\mathbf{p}}(t')}{\partial t'}$ in Eqs.~\eqref{saddle_point_equations} and \eqref{SFA_saddle_point_equations}. In the first order of perturbation w.r.t. $G_C$, all deviations from the SFA saddle-point solutions are proportional to $G_C$: $\Delta\theta_s' \sim \mathcal{O}(G_C)$, $\Delta\phi_s' \sim \mathcal{O}(G_C)$, and $\Delta t_a' \sim \mathcal{O}(G_C)$. In our analysis we shall consistently keep only terms of the order $\sim \mathcal{O}(G_C)$ and therefore, only the SFA saddle points can enter the argument of $G_C(\mathbf{p},T; \mathbf{a},t')$ and the SFA action. Indeed using Eq.~\eqref{SFA_saddle_point_equations} we obtain
	\begin{align}
		G_C(\mathbf{p},T; \mathbf{a}, t') &= G_C\left(\mathbf{p},T; \mathbf{r}_{s}^{\prime(0)}, t_a^{\prime(0)}\right) + \mathcal{O}(G_C^2), \label{expansion1}\\
		S^{\text{SFA}}(\mathbf{p}, T; \mathbf{a}, t') &= S^{\text{SFA}}\left(\mathbf{p}, T; \mathbf{r}_{s}^{\prime(0)},t_a^{\prime(0)}\right) + \mathcal{O}(G_C^2). \label{expansion2}
	\end{align}
However, the corrected saddle-point solution for time will contribute to the pre-exponential factor $P(\theta', \phi', t')$, since $\frac{\partial P(\theta',\phi',t')}{\partial t'} \neq 0$. It is straightforward to show that $\theta_s^{\prime(0)} = \theta_v(t')$ and $\phi_s^{\prime(0)} = \phi_v(t')$, where $\theta_v(t')$ and $\phi_v(t')$ describe the direction of electron velocity at time $t'$.

\subsection{Integration over the surface of the sphere} \label{surface_integral}

Because of the large value of the action, only a vicinity of saddle points contributes to the integral. We do a Taylor expansion of $G_C$ around points $\theta_s^{\prime(0)}$, $\phi_s^{\prime(0)}$, and $t_a^{\prime(0)}$ [only the saddle point in time is affected by the boundary term $av_{\mathbf{p}}(t')$] up to quadratic terms
	\begin{equation}
		\begin{split}
			&G_C(\mathbf{p},T;\mathbf{a}, t') = G_C\left(\mathbf{p},T; \mathbf{r}_s^{\prime(0)},t_a^{\prime(0)}\right) + \left(\mathbf{a}-\mathbf{r}_s^{\prime(0)}\right)\cdot\nabla G_C\left(\mathbf{p},T;\mathbf{r}_s^{\prime(0)},t_a^{\prime(0)}\right)
			\\&+ \left(t'-t_a^{\prime(0)}\right)\partial_{t'}G_C\left(\mathbf{p},T; \mathbf{r}_s^{\prime(0)}, t_a^{\prime(0)}\right) + \frac{1}{2}\left(t'-t_a^{\prime(0)}\right)^2\partial_{t'}^2G_C\left(\mathbf{p},T;\mathbf{r}_s^{\prime(0)}, t_a^{\prime(0)}\right). \label{quadratic_expansion}
		\end{split}
	\end{equation}
The term involving the mixed derivative $\left(t - t_{a}^{\prime(0)}\right)\left(\mathbf{a} - \mathbf{r}_s^{\prime(0)}\right)\cdot\nabla\partial_{t'}G_C\left(\mathbf{p},T; \mathbf{r}_s^{\prime(0)},t_a^{\prime(0)}\right) \propto \mathcal{O}(G_C^2)$ is omitted from Eq.~\eqref{quadratic_expansion}, as a higher order correction, since $\nabla G_C\left(\mathbf{p}, T; \mathbf{r}_s^{\prime(0)}, t_a^{\prime(0)}\right)$ is multiplied to $t' - t_a^{\prime(0)} \propto \mathcal{O}(G_C)$. The term involving second derivatives w.r.t. spatial coordinates on the surface of the sphere $\frac{1}{2}\left(\mathbf{a}-\mathbf{r}_s^{\prime(0)}\right)^2\Delta G_C\left(\mathbf{p},T;\mathbf{r}_s^{\prime(0)}, t_a^{\prime(0)}\right)$ is equal to 0 for Coulomb potential, since $\Delta U(r)=\delta(r)$ and the argument of $U$ in $G_C\left(\mathbf{p},T; \mathbf{r}_s^{\prime(0)}, t_a^{\prime(0)}\right)$ is trajectory starting at the surface and propagating outside of the sphere; this trajectory never reaches the origin. Note that $\nabla G_C = -\Delta \mathbf{p}$, where $\Delta \mathbf{p}$ is the modification of the canonical momentum arising due to electron interaction with the long-range potential of the core
	\begin{equation}
		\Delta\mathbf{p}(t',T) \equiv -\nabla G_C = -\int_{t'}^{T}d\tau\,\frac{U'}{\left\|\mathbf{r}' + \int_{t'}^{\tau}d\zeta\,\mathbf{v}_{\mathbf{p}}(\zeta)\right\|}\left[\mathbf{r}' + \int_{t'}^{\tau}d\zeta\,\mathbf{v}_{\mathbf{p}}(\zeta)\right], \label{momentum_shifts}
	\end{equation}
where $U'$ represents a derivative of $U$ w.r.t. its argument. Indeed, $\Delta\mathbf{p}(t',T)$ is given by the integral from the force $\mathbf{F}=-\nabla U$, calculated along the electron trajectory
 \begin{equation}
		\Delta\mathbf{p}(t',T) \equiv \int_{t'}^{T}d\tau\,\mathbf{F}\left[\mathbf{r}' + \int_{t'}^{\tau}d\zeta\,\mathbf{v}_{\mathbf{p}}(\zeta)\right]. \label{momentum_shifts1}
	\end{equation}
It is convenient to rewrite the time derivative of $G_C(\mathbf{p},T; \mathbf{a}, t')$ as \cite{olga2008}
	\begin{equation}
		\begin{split}
			\partial_{t'}G_C\left(\mathbf{p},T; \mathbf{r}_s^{\prime(0)},t_a^{\prime(0)}\right)
&= -\mathbf{v}_{\mathbf{p}}\left(t_a^{\prime(0)}\right)\cdot\nabla G_C\left(\mathbf{p},T; \mathbf{r}_s^{\prime(0)},t_a^{\prime(0)}\right) - U(a) \\ &= -\left[-\mathbf{v}_{\mathbf{p}}\cdot\Delta\mathbf{p}\left(t_a^{\prime(0)},T\right) + U(a)\right]. \label{HJ_equation}
		\end{split}
	\end{equation}
Substituting Eq.~\eqref{quadratic_expansion} into the expression for $a_{\mathbf{p}}(T)$, and evaluating the integral over $\phi'$ and $\theta'$ exactly (see Appendix~\ref{subapp:wave function}), we obtain
	\begin{equation}
		\begin{split}
			a_{\mathbf{p}}(T) &= N_{\ell m}(-i)^{\ell}(-1)^m\varphi_{\kappa\ell}(a)\frac{2i\kappa a^2}{\sqrt{2\pi}}e^{-iG_C\left(\mathbf{p},T; \mathbf{r}_s^{\prime(0)},t_a^{\prime(0)}\right)}
\int_{t_0}^{T}dt'\,e^{-\frac{i}{2}\int_{t'}^{T}d\tau\,v_{\mathbf{p}}^2(\tau) + i\kappa^2(t'-t_0)/2}\\&e^{-i\left(t'-t_s^{\prime(0)}\right)\partial_{t'}G_C\left(\mathbf{p},T; \mathbf{r}_s^{\prime(0)}, t_s^{\prime(0)}\right)}e^{im\phi_v^c(t')}P_{\ell}^m\left(\frac{p_z^c}{v_{\mathbf{p}^c}(t')}\right)j_{\ell}\left(av_{\mathbf{p}}(t')\right), \label{a_pTdef4}
		\end{split}
	\end{equation}
where $\phi_v^c(t')$ is the tunneling angle,
	\begin{equation}
		\tan\phi_v^c(t')=\frac{p_y - \Delta p_y(t',T) + A_y(t')}{p_x - \Delta p_x(t',T) + A_x(t')}, \label{tunneling_angle}
	\end{equation}
and $\mathbf{p}^c = \mathbf{p} - \Delta\mathbf{p}(t',T)$ is the Coulomb-shifted momentum at time $t'$ corresponding to the asymptotic momentum $\mathbf{p}$ registered at the detector at time $T$,
	\begin{equation}
		\Delta\mathbf{p}(t',T)=-\nabla G_C\left(\mathbf{p},T; \mathbf{r}_s^{\prime(0)},t'\right). \label{Dp_def}
	\end{equation}
Note that the tunneling angle is complex. It signifies the sensitivity of strong-field ionization to the sense of rotation of the electron in the initial state \cite{barth2011, barth2013}.

\subsection{Integration over time}

We are now left with the integral over $t'$. We use the saddle-point method to evaluate this integral. The saddle-point equation for $t'$ is
	\begin{equation}
		\frac{\partial S}{\partial t'} = \frac{\partial S^{\text{SFA}}}{\partial t'} + \frac{\partial G_C}{\partial t'} + \left(t'-t_a^{\prime(0)}\right)\frac{\partial^2 G_C}{\partial t'^2} + av_{\mathbf{p}}'(t')=0, \label{saddle_t}
	\end{equation}
where the last term, as discussed in Sec.~\ref{subsection:boundary_matching}, comes from $j_{\ell}\left(av_{\mathbf{p}}(t')\right)$. To solve this equation, we expand the derivative of the SFA action in Eq.~\eqref{saddle_t} up to quadratic terms w.r.t. $\Delta t_a^{\prime(0)}$ and take into account that $\partial_{t'}S_a^{\text{SFA}}\left(t_a^{\prime(0)}\right)=0$, yielding
	\begin{equation}
		\Delta t_a^{\prime(0)} = -\frac{\partial_{t'}G_C\left(t_a^{\prime(0)}\right)}{\partial_{t'}^2S_a^{\text{SFA}}\left(t_a^{\prime(0)}\right) + \partial_{t'}^2G_C\left(t_a^{\prime(0)}\right)} \simeq -\frac{\partial_{t'}G_C\left(t_a^{\prime(0)}\right)}{\partial_{t'}^2S_a^{\text{SFA}}\left(t_a^{\prime(0)}\right)} \simeq -\frac{\partial_{t'}G_C\left(t_a^{\prime(0)}\right)}{\partial_{t'}^2S_a^{\text{SFA}}\left(t_s^{\prime(0)}\right)}. \label{dt_a}
	\end{equation}
Here we have used that $t_s^{\prime(0)} - t_a^{\prime(0)} = -ia/\kappa \ll t_s^{\prime(0)}$ \cite{lisa2012}. We have omitted terms of order of $G_C$ in the denominator in the last term in Eq.~\eqref{dt_a}, since the terms of the first order of $G_C$ are already included in the denominator. Taking into account Eq.~\eqref{HJ_equation} and
	\begin{equation}
		\partial_{t'}^2S_a^{\text{SFA}}\left(t_s^{\prime(0)}\right) = \mathbf{E}\left(t_s^{\prime(0)}\right)\cdot\mathbf{v}_{\mathbf{p}}
\left(t_s^{\prime(0)}\right),
	\end{equation}
we obtain for the time
	\begin{equation}
		\Delta t_a^{\prime(0)} = \frac{-\mathbf{v}_{\mathbf{p}}\left(t_s^{\prime(0)}\right)\cdot\Delta\mathbf{p}\left(t_a^{\prime(0)},T\right) + U(a)}{\mathbf{E}\left(t_s^{\prime(0)}\right)\cdot\mathbf{v}_{\mathbf{p}}\left(t_s^{\prime(0)}\right)}. \label{dt_afin}
	\end{equation}
Equations~\eqref{expansion1} and \eqref{expansion2} suggest that Coulomb corrections to the ionization time do not affect the exponent of the ionization amplitude, but they contribute to the prefactor, further modifying the tunneling angle:
	\begin{equation}
		\tan\phi_v^c\left(t_a^{\prime(0)} + \Delta t_a^{\prime(0)}\right)=\frac{p_y - \Delta p_y\left(t_a^{\prime(0)}+\Delta t_a^{\prime(0)},T\right) + A_y\left(t_a^{\prime(0)} + \Delta t_a^{\prime(0)}\right)}{p_x - \Delta p_x\left(t_a^{\prime(0)}+\Delta t_a^{\prime(0)},T\right) + A_x\left(t_a^{\prime(0)} + \Delta t_a^{\prime(0)}\right)} \label{tunneling_angle_t_a(1)}
	\end{equation}
Up to first order terms w.r.t. to $G_C$, $a_{\mathbf{p}}(T)$ is
	\begin{equation}
		\begin{split}
			a_{\mathbf{p}}(T) &= i\kappa a^2\varphi_{\kappa\ell}(a)N_{\ell m}\sqrt{\frac{1}{S''\left(t_a^{\prime(1)}\right)}}e^{-\frac{i}{2}\int_{t_s^{\prime(0)}}^{T}d\tau\,v_{\mathbf{p}}^2(\tau) + i\kappa^2t_s^{\prime(0)}/2 - iG_C\left(\mathbf{p},T; \mathbf{r}_s^{\prime(0)},t_a^{\prime(0)}\right)}\\&
e^{im\phi_v^c\left(t_a^{\prime(1)}\right)}P_{\ell}^m\left(\frac{p_z^c}{v_{\mathbf{p}^c}\left(t_a^{\prime(1)}\right)}\right)
j_{\ell}\left(av_{\mathbf{p}}\left(t_a^{\prime(1)}\right)\right). \label{a_pT}
		\end{split}
	\end{equation}

\subsection{Boundary matching} \label{subsection:boundary_matching}

We now consider the elimination of boundary dependence in the results for transition amplitude.
In the long pulse, due to cylindrical symmetry of the problem, the result does not depend on the position of the detector in the polarization plane $x,y$. Thus, without loss of generality we consider the electron registered at the detector placed in the positive direction of the $x$ axis, i.e., the electron momentum at the detector $p_y = 0$.

\subsubsection{Complex momentum $\Delta\mathbf{p}\left(t_a^{\prime(0)},T\right)$ at the boundary} \label{subsubsection:boundary_matching_momentum}

To perform boundary matching in Eq.~\eqref{dt_afin} and Eq.~\eqref{tunneling_angle_t_a(1)}, we need to evaluate the  momentum $\Delta\mathbf{p}\left(t_a^{\prime(0)},T\right)$ at the boundary
	\begin{equation}
		\Delta\mathbf{p}\left(t_a^{\prime(0)},T\right)=-\int_{t_a^{\prime(0)}}^{T}d\tau\,\nabla U\left(\mathbf{r}_s^{\prime(0)}+\int_{t_a^{\prime(0)}}^{\tau}d\zeta\,\mathbf{v}_{\mathbf{p}}(\zeta)\right). \label{Dp_complexa}
	\end{equation}
Note that $\Delta\mathbf{p}\left(t_a^{\prime(0)},T\right)$ is a function of the final momentum $\mathbf{p}$. In this section we consider only  $\mathbf{p}=\mathbf{p}_{\text{opt}}$, corresponding to the momentum at which the probability is maximal, since it is sufficient to calculate the ionization rates. The photoelectron spectra will be considered in our subsequent publication \cite{lisa2013}. In the polarization plane the optimal momentum $\mathbf{p}_{\text{opt}} = (p_{\text{opt}}\cos\phi_p, p_{\text{opt}}\sin\phi_p)$ is given by the radial momentum
	\begin{equation}
		p_{\text{opt}} = A_0\sqrt{1+\gamma^2}\sqrt{\frac{1-\zeta_0}{1+\zeta_0}}, \label{optimal_momentum}
	\end{equation}
for any angle $\phi_p$. The parameter $0 \leq \zeta_0 \leq 1$ satisfies the equation $\sqrt{\frac{\zeta_0^2+\gamma^2}{1+\gamma^2}} = \tanh\frac{1}{1-\zeta_0}\sqrt{\frac{\zeta_0^2+\gamma^2}{1+\gamma^2}}$ \cite{ppt1966, barth2011, barth2013}. Note that $\zeta_0 \simeq \gamma^{2}/3$ for $\gamma \ll 1$, and $\zeta_0 \simeq 1-1/\ln\gamma$ for $\gamma \gg 1$ \cite{ppt1966}.
An alternative expression for $p_{\text{opt}}$ is
	\begin{equation}
		p_{\text{opt}} = A_0\frac{\sinh\omega\tau_i^{\prime(0)}}{\omega\tau_i^{\prime(0)}}, \label{optimal_momentum1}
	\end{equation}
where $\tau_i^{\prime(0)} = \Im\left[t_s^{\prime(0)}\right]$, is the imaginary part of the saddle-point solution for time, also known as the ``tunneling time." The advantage of the second expression is that it provides a compact connection between the optimal momentum and the tunneling time, however, one has to keep in mind that in a circular field $\tau_i^{\prime(0)}$ depends on the final radial momentum $p$ \cite{ppt1966,barth2011,barth2013}:
	\begin{align}
		\omega\tau_i^{\prime(0)} = \operatorname{\cosh^{-1}}\eta,\quad
		\eta(\mathbf{p}) = \frac{A_0}{2p_{\rho}}\left[\left(\frac{p}{A_0}\right)^2 + \gamma^2 + 1\right], \label{eta}
	\end{align}
and thus, in Eq.~\eqref{optimal_momentum1}, $\tau_i^{\prime(0)}$ depends on $p_{\text{opt}}$ itself.

Since the time $t_a^{\prime(0)}$ is complex, the momentum $\Delta\mathbf{p}\left(t_a^{\prime(0)},T\right)$ will also be complex:
	\begin{align}
		\Delta p_y\left(t_a^{\prime(0)},T\right) &= \Delta p_y^{\text{re}}(a) + i\Delta p_y^{\text{im}}(a), \label{Dpy_complex}\\
		\Delta p_x\left(t_a^{\prime(0)},T\right) &= \Delta p_x^{\text{re}}(a) + i\Delta p_x^{\text{im}}(a). \label{Dpx_complex}
	\end{align}
After some algebra (see Appendix~\ref{app:momenta}) we obtain:
	\begin{gather}
		\Delta p_y^{\text{im}}(a) \simeq \mathcal{O}\left(\frac{1}{\kappa a}\right) \to 0, \label{Dpy_imagfin}\\
		\Delta p_x^{\text{re}}(a) = \Delta p_x^{\text{re}}, \quad \Delta p_y^{\text{re}}(a) = \Delta p_y^{\text{re}}, \label{Dpy_realfin}\\
		\Delta p_x^{\text{im}}(a) \simeq \mathcal{O}\left(\frac{1}{\kappa a}\right) \to 0, \label{Dpx_imagfin}
	\end{gather}
where the boundary-independent momentum is
	\begin{equation}
		\Delta\mathbf{p}^{\text{re}} = -\int_{\Re\left[t_s^{\prime(0)}\right]}^{T}d\tau\,\nabla U\left(\mathbf{r}_e^{\prime(0)}+\int_{\Re\left[t_s^{\prime(0)}\right]}^{\tau}d\zeta\,\mathbf{v}_{\mathbf{p}_{\text{opt}}}(\zeta)\right), \label{Dp_realfin}
	\end{equation}
and the coordinate $\mathbf{r}_e^{\prime(0)}$, known as the coordinate of exit from the tunneling barrier, is defined as
	\begin{equation}
		\mathbf{r}_e^{\prime(0)}=\int_{t_s^{\prime(0)}}^{\Re\left[t_s^{\prime(0)}\right]}d\zeta\,\mathbf{v}_{\mathbf{p}_{\text{opt}}}(\zeta). \label{opt_coord_exit}
	\end{equation}

\subsubsection{Boundary matching for $\Delta t_a^{\prime(0)}$ in Eq.~\eqref{dt_afin} and for the tunneling angle $\tan\phi_v^c\left(t_a^{\prime(0)} + \Delta t_a^{\prime(0)}\right)$ in Eq.~\eqref{tunneling_angle_t_s(1)}} \label{subsubsection:boundary_matching_time}

Substituting Eqs.~\eqref{Dpy_imagfin}, \eqref{Dpy_realfin}, \eqref{Dpx_imagfin}, and \eqref{Dp_realfin} for $\Delta\mathbf{p}\left(t_a^{\prime(0)},T\right)$ from the previous section into Eq.~\eqref{dt_afin} and taking into account that in our geometry
	\begin{align}
		E_y\big(t_s^{\prime(0)}\big) &= E_y^{\text{re}}, \quad E_y^{\text{re}} = E_0\cosh\omega\tau_i^{\prime(0)}, \label{Ey_in}\\
		E_x\big(t_s^{\prime(0)}\big) &= i E_x^{\text{im}}, \quad E_x^{\text{im}} = -E_0\sinh\omega\tau_i^{\prime(0)}, \label{Ex_in}\\
		v_y\big(t_s^{\prime(0)}\big) &= iv_y^{\text{im}}, \quad v_y^{\text{im}} = -A_0\sinh\omega\tau_i^{\prime(0)}, \label{vy_in}
	\end{align}
	\begin{equation}
		v_x\big(t_s^{\prime(0)}\big) = v_x^{\text{re}}, \quad v_x^{\text{re}} = p_{\text{opt}} - A_0\cosh\omega\tau_i^{\prime(0)} = \frac{a_0}{\tau_i^{\prime(0)}}\left(\sinh\omega\tau_i^{\prime(0)} - \omega\tau\cosh\omega\tau_i^{\prime(0)}\right), \label{vx_in}
	\end{equation}
where $a_0 = E_0/\omega^2$ is the electron oscillation amplitude, yielding
	\begin{equation}
		\mathbf{E}\left(t_s^{\prime(0)}\right)\cdot\mathbf{v}_{\mathbf{p}_{\text{opt}}}
\left(t_s^{\prime(0)}\right) = ip_{\text{opt}}E_0\sinh\omega\tau_i^{\prime(0)} = i v_y^{\text{im}}p_{\text{opt}}\omega,
	\end{equation}
we obtain
	\begin{align}
		\Re\left[\Delta t_a^{\prime(0)}\right] &= \frac{-v_{x}^{\text{re}}\Delta p_{x}^{\text{im}}/v_{y}^{\text{im}} - \Delta p_{y}^{\text{re}}}{p_{\text{opt}}\omega},\\
		\Im\left[\Delta t_a^{\prime(0)}\right] &= \frac{v_x^{\text{re}}\Delta p_x^{\text{re}}/v_y^{\text{im}} - U(a)/v_y^{\text{im}} - \Delta p_{y}^{\text{im}}}{p_{\text{opt}}\omega}.
	\end{align}
Since $U(a)/v_y^{\text{im}} \simeq \Delta p_y^{\text{im}} \simeq \mathcal{O}\left(\frac{1}{\kappa a}\right) \to 0$ (see Appendix~\ref{app:momenta}), $\Delta p_x^{\text{im}} \simeq \mathcal{O}\left(\frac{1}{\kappa a}\right) \to 0$, we obtain a boundary-independent correction to the real ionization time $\Re\left[\Delta t_s^{\prime(0)}\right] = \Re\left[\Delta t_a^{\prime(0)}\right]$ and imaginary ionization time $\Im\left[\Delta t_s^{\prime(0)}\right] = \Im\left[\Delta t_a^{\prime(0)}\right]$:
	\begin{align}
		\Re\left[\Delta t_s^{\prime(0)}\right] &= -\frac{\Delta p_y^{\text{re}}}{p_{\text{opt}}\omega} = -\frac{\Delta p_y^{\text{re}}}{E_0}\frac{\omega\tau_i^{\prime(0)}}{\sinh\omega\tau_i^{\prime(0)}}, \label{re_time}\\
		\Im\left[\Delta t_s^{\prime(0)}\right] &= \frac{\Delta p_x^{\text{re}}v_x^{\text{re}}}{v_y^{\text{im}}p_{\text{opt}}\omega} = -\frac{\Delta p_x^{\text{re}}}{E_0}\frac{\sinh\omega\tau_i^{\prime(0)} - \omega\tau\cosh\omega\tau_i^{\prime(0)}}{\sinh^2\omega\tau_i^{\prime(0)}}, \label{im_time}
	\end{align}
where the subscript ``$s$" denotes that the results for corrections to the SFA saddle point $t_a^{\prime(0)}$ are now independent of the boundary $r' = a$. Thus we can write the saddle point as
	\begin{equation}
		t_s^{\prime(1)} = t_s^{\prime(0)} + \Delta t_s^{\prime(0)}. \label{full_time}
	\end{equation}
Matching for the tunneling angle is now trivial, since all variables entering Eq.~\eqref{tunneling_angle_t_a(1)} are now proved to be boundary independent:
	\begin{equation}
		\tan\phi_v^c\left(t_a^{\prime(0)} + \Delta t_a^{\prime(0)}\right) = \frac{v_y\left(t_s^{\prime(0)}\right) - \Delta p_y - \Delta t_s^{\prime(0)}E_y}{v_x\left(t_s^{\prime(0)}\right) - \Delta p_x - \Delta t_s^{\prime(0)}E_x}. \label{tunneling_angle_t_s(1)}
	\end{equation}

\subsubsection{Boundary matching of the remaining terms in Eq.~\eqref{a_pT}} \label{subsubsection:boundary_matching_rest}

We first establish the connection (in Appendix~\ref{subapp:additional}):
	\begin{equation}
		j_{\ell}\left(av_{\mathbf{p}^c}\left(t_a^{\prime(1)}\right)\right)e^{-i\mathbf{r}_s^{\prime(0)}\cdot\Delta\mathbf{p}} = j_{\ell}\left(av_{\mathbf{p}}\left(t_s^{\prime(0)}\right)\right) \label{j_l_matching}
	\end{equation}
Next, we consider matching of the EVA phase to the bound wave function. We follow the approach used in \cite{lisa2012} for a linearly polarized field:
	\begin{equation}
		B(a) = \kappa a^2\varphi_{\kappa\ell}(a)e^{iG_C\left(\mathbf{r}_s^{\prime(0)},t_a^{\prime(0)},T\right)}
j_{\ell}\left(av_{\mathbf{p}}\left(t_s^{\prime(0)}\right)\right)
	\end{equation}
The asymptotic bounded wave function in a Coulomb potential is
	\begin{equation}
		\varphi_{\kappa\ell}(r) = C_{\kappa l}\kappa^{3/2}\frac{e^{-\kappa r}}{\kappa r}(\kappa r)^{Q/\kappa}.
	\end{equation}
Furthermore, at $r' = a$, we can write $(\kappa a)^{Q/\kappa}$ as
	\begin{equation}
		(\kappa a)^{Q/\kappa} = e^{Q/\kappa\int_{1/\kappa}^{a}\frac{d\chi}{\chi}} = e^{-i\int_{t_{\kappa}^{\prime(0)}}^{t_a^{\prime(0)}}d\tau\,U\left(\int_{t_s^{\prime(0)}}^{\tau}d\zeta\,i\kappa\right)} = e^{-i\int_{t_{\kappa}^{\prime(0)}}^{t_a^{\prime(0)}}d\tau\,U\left(\int_{t_s^{\prime(0)}}^{\tau}d\zeta\,\mathbf{v}(\zeta)\right)}, \label{bound}
	\end{equation}
where $t_{\kappa}^{\prime(0)} = t_s^{\prime(0)} - i/\kappa^2$ and $t_a^{\prime(0)} = t_s^{\prime(0)} - ia/\kappa$. The second equality follows from the fact that between $t_{\kappa}^{\prime(0)}$ and $t_a^{\prime(0)}$, the velocity of the electron remains almost constant, while the third equality holds because, finally, we will be using the modulus of the vector, and under the approximation
	\begin{equation}
		\left\|\int_{t_s^{\prime(0)}}^{\tau}d\zeta\,\mathbf{v}(\zeta)\right\|\approx\|\mathbf{v}(t_s^{\prime(0)})\|\left(\tau-t_s^{\prime(0)}\right) = i\kappa\left(\tau-t_s^{\prime(0)}\right) = \int_{t_s^{\prime(0)}}^{\tau}d\zeta\,i\kappa.
	\end{equation}
The term given by Eq.~\eqref{bound} can now be matched with the Coulomb phase \allowbreak term $G_C\left(\mathbf{p},T; \mathbf{r}_s^{\prime(0)},t_a^{\prime(0)}\right)$:
	\begin{equation}
		G_C\left(\mathbf{p}, T; \mathbf{r}_s^{\prime(0)}, t_a^{\prime(0)}\right) = \int_{t_a^{\prime(0)}}^{T}d\tau\,U\left(\mathbf{r}_s^{\prime(0)} + \int_{t_a^{\prime(0)}}^{\tau}d\zeta\,\mathbf{v}_{\mathbf{p}}(\zeta)\right) = \int_{t_a^{\prime(0)}}^{T}d\tau\,U\left(\int_{t_s^{\prime(0)}}^{\tau}d\zeta\,\mathbf{v}_{\mathbf{p}}(\zeta)\right)
	\end{equation}
to yield, after using the large-argument approximation for $j_{\ell}\left(av_{\mathbf{p}}\left(t_s^{\prime(0)}\right)\right)$,
	\begin{equation}
		\begin{split}
			B(a) &= -iC_{\kappa l}\sqrt{\kappa}e^{-\kappa a}e^{-i\int_{t_{\kappa}^{\prime(0)}}^{t_a^{\prime(0)}}d\tau\,
U\left(\int_{t_s^{\prime(0)}}^{\tau}d\zeta\,\mathbf{v}_{\mathbf{p}}(\zeta)\right)}e^{i\int_{T}^{t_a^{\prime(0)}}U
\left(\int_{t_s^{\prime(0)}}^{\tau}d\zeta\,\mathbf{v}_{\mathbf{p}}(\zeta)\right)}\\&\left(e^{-\kappa a-i(\ell+1)\pi/2} + e^{\kappa a+i(\ell+1)\pi/2}\right)\\
			&= i^{\ell}C_{\kappa l}\sqrt{\kappa}e^{i\int_{T}^{t_{\kappa}^{\prime(0)}}d\tau\,
U\left(\int_{t_s^{\prime(0)}}^{\tau}d\zeta\,\mathbf{v}(\zeta)\right)}
\left(1+(-1)^{\ell+1}e^{-2\kappa a}\right),
		\end{split}
	\end{equation}
which ensures boundary matching for all transition rates and amplitudes.

After boundary matching the final expression for the ionization amplitude is independent of $a$:
	\begin{equation}
		\begin{split}
			a_{\mathbf{p}}(T) &= (-1)^mC_{\kappa\ell}N_{\ell m}\sqrt{\frac{\kappa}{S''\left(t_s^{\prime(1)}\right)}}e^{-\frac{i}{2}\int_{t_s^{\prime(0)}}^{T}d\tau\,v_{\mathbf{p}}^2(\tau) + i\frac{\kappa^2}{2}t_s^{\prime(0)}}e^{-i\int_{t_{\kappa}^{\prime(0)}}^{T}d\tau\,U\left(\int_{t_s^{\prime(0)}}^{\tau}d\zeta\,\mathbf{v}_{\mathbf{p}}(\zeta)\right)}\\&P_{\ell}^m\left(\frac{p_z^c}{v_{\mathbf{p}^c}\left(t_s^{\prime(1)}\right)}\right)e^{im\phi_v^c\left(t_s^{\prime(1)}\right)}. \label{ion_amp}
		\end{split}
	\end{equation}
Here $t_s^{\prime(1)}$ is given by Eq.~\eqref{full_time}. Equation~\eqref{ion_amp} corresponds to a ``single" ionization event (ionization amplitude formed after one laser cycle), since only one saddle point is included.

\subsection{Ionization rate}


To calculate the ionization rate we integrate the ionization amplitude (corresponding to a single ionization event) over all momenta using the saddle-point method and divide by the period of the laser field:
	\begin{equation}
		w = \frac{\omega}{2\pi}\int_{}^{}d\mathbf{p}\,\left|a_{\mathbf{p}}(T)\right|^2. \label{ion_rate}
	\end{equation}
In our forthcoming publication \cite{lisa2013} we will analyze the accuracy of saddle point approximation in Eq.~\eqref{ion_rate}. As follows from Eq.~\eqref{ion_amp}, the ionization amplitude $a_{\mathbf{p}}(T)$ can be written in the form $a_{\mathbf{p}}(T) = P_{\mathbf{p}}e^{-iF_{\mathbf{p}}}$ and the integral Eq.~\eqref{ion_rate} can be calculated using the saddle-point method.

The saddle-point equation
	\begin{equation}
		\nabla_{\mathbf{p}} 2 \Im[F_{\mathbf{p}}] = 2\nabla_{\mathbf{p}}\Im[S^{\text{SFA}}_{\mathbf{p}}] + 2\nabla_{\mathbf{p}}\left[\Im[ F_{\mathbf{p}}]-\Im[S^{\text{SFA}}_{\mathbf{p}}]\right] = 0 \label{saddle_p}
   \end{equation}
can again be solved iteratively, since the second term is small by construction. The optimal  momentum in SFA solves the equation
	\begin{equation}
		 2\nabla_{\mathbf{p}}\Im[S^{\text{SFA}}_{\mathbf{p}}] = 0 \label{SFA_saddle_p}
	\end{equation}
and is given by Eqs.~\eqref{optimal_momentum} and \eqref{optimal_momentum1}
Since the correction to $p_{\text{opt}}$ are obtained from Eq.~\eqref{saddle_p}, they will contribute to the ionization rate in the second order w.r.t. $G_C$. We keep only terms first order in $G_C$ and therefore these corrections are irrelevant and the saddle point for the momentum integral in the ionization rate is given by the optimal momentum, Eq.~\eqref{optimal_momentum}. We neglect here small corrections arising from substituting the pre-exponential factor $S_{\mathbf{p}}''\left(t_s^{\prime(1)}\right)$ with $S_{\mathbf{p}}''\left(t_s^{\prime(0)}\right)$ in Eq.~\eqref{ion_amp}. $S''$ denotes the derivative of the action w.r.t. the radial momentum. Finally, using the saddle-point method for the radial integral and taking into account that integration over $\phi_p$ yields $2\pi$, we obtain the expression for the ionization rate,
	\begin{equation}
		\begin{split}
			w_{\text{opt}} = |C_{\kappa\ell}|^2|N_{\ell m}|^2\frac{\gamma}{\sqrt{\eta^2-1}}\sqrt{\frac{\pi}{\Im[S''(p_{\rho,\text{opt}})]}}e^{-\frac{4n_0}{1-\zeta_0^2}\sqrt{\frac{\zeta_0^2+\gamma^2}{1+\gamma^2}}}e^{2W_{C1}+2W_{C2}}\\\left|P_{\ell}^m\left(\frac{-\Delta k_z}{v_{\mathbf{p}^c}\left(t_s^{\prime(1)}\right)}\right)\right|^2e^{-2m\Im\left[\phi_v^c\left(t_s^{\prime(1)}\right)\right]}, \label{ion_rate_opt}
		\end{split}
	\end{equation}
where
	\begin{align}
		\Im[S''(p_{\text{opt}})] &= \frac{2\zeta_0^2 + \gamma^2(1+\zeta_0^2)}{\omega (1-\zeta_0)\sqrt{(\zeta_0^2+\gamma^2)(1+\gamma^2)}},\\
		W_{C1} &= -\int_{0}^{\tau_{\kappa}^{\prime(0)}}d\tau\,\Re\left[U\left( \int_{t_s^{\prime(0)}}^{\tau}d\zeta\,\mathbf{v}_{\mathbf{p}}(\zeta)\right)\right],\\
		W_{C2} &= \int_{t_i^{\prime(0)}}^{T}d\tau\Im\left[U\left(\int_{t_s^{\prime(0)}}^{\tau}d\zeta\,\mathbf{v}_{\mathbf{p}}(\zeta)\right)\right].
	\end{align}
$W_{C1}$ is a well-known adiabatic Coulomb correction, evaluated under the barrier along the optimal trajectory \cite{ppt1966,popruzhenko2008}.
Analysis of Eq.~\eqref{ion_rate_opt} shows that nonadiabatic Coulomb effects modify the ionization dynamics in several ways. New effects arising from our analysis include modification of (i) ionization times, (ii) initial conditions for electron continuum dynamics, and (iii) the ``tunneling angle".

We discuss  these  Coulomb effects in detail in the next section.
We show that  Coulomb effects modify (i) calibration of the attoclock \cite{eckle2008, eckle2008-2, pfeiffer2011, pfeiffer2012} in the angular streaking method, and (ii)  the ratio of ionization rates from $p^{-}$ and $p^{+}$ orbitals obtained for short-range potentials in \cite{barth2011}. The photoelectron spectra will be considered in our subsequent publication \cite{lisa2013}, where we will include the effects of $W_{C2}$, the result of interaction of the long-range potential with the electron in the continuum and depart from the saddle point approximation in Eq.~\eqref{ion_rate}.

\section{Physical picture of ionization in long range potentials} \label{section:phys_pic}

In circularly polarized fields, the electron liberated at different times will be ``directed" by the laser field into different angles. This idea is called ``angular streaking"  and the corresponding ``time-to-angle" mapping is unique for nearly single-cycle pulses with a stable carrier-envelope phase, underlying the idea of the attoclock \cite{eckle2008,eckle2008-2,pfeiffer2011,pfeiffer2012}. The angular streaking principle makes single and double ionization in circularly polarized strong laser fields a sensitive probe of the attosecond dynamics \cite{eckle2008,eckle2008-2,pfeiffer2011,pfeiffer2012,akagi2009,fleischer2011}.

However, reconstruction of this dynamics requires the calibration of the attoclock, i.e., establishing the mapping between the direction of the laser polarization vector at the time of ionization and the direction of the electron momentum at the detector. When one strives to achieve the accuracy of, say, 10 as, using an 800-nm carrier as a clock, one needs to know this mapping with an accuracy of about $1^{\circ}$.
\begin{figure}
	\subfigure[]{\includegraphics[scale=0.4]{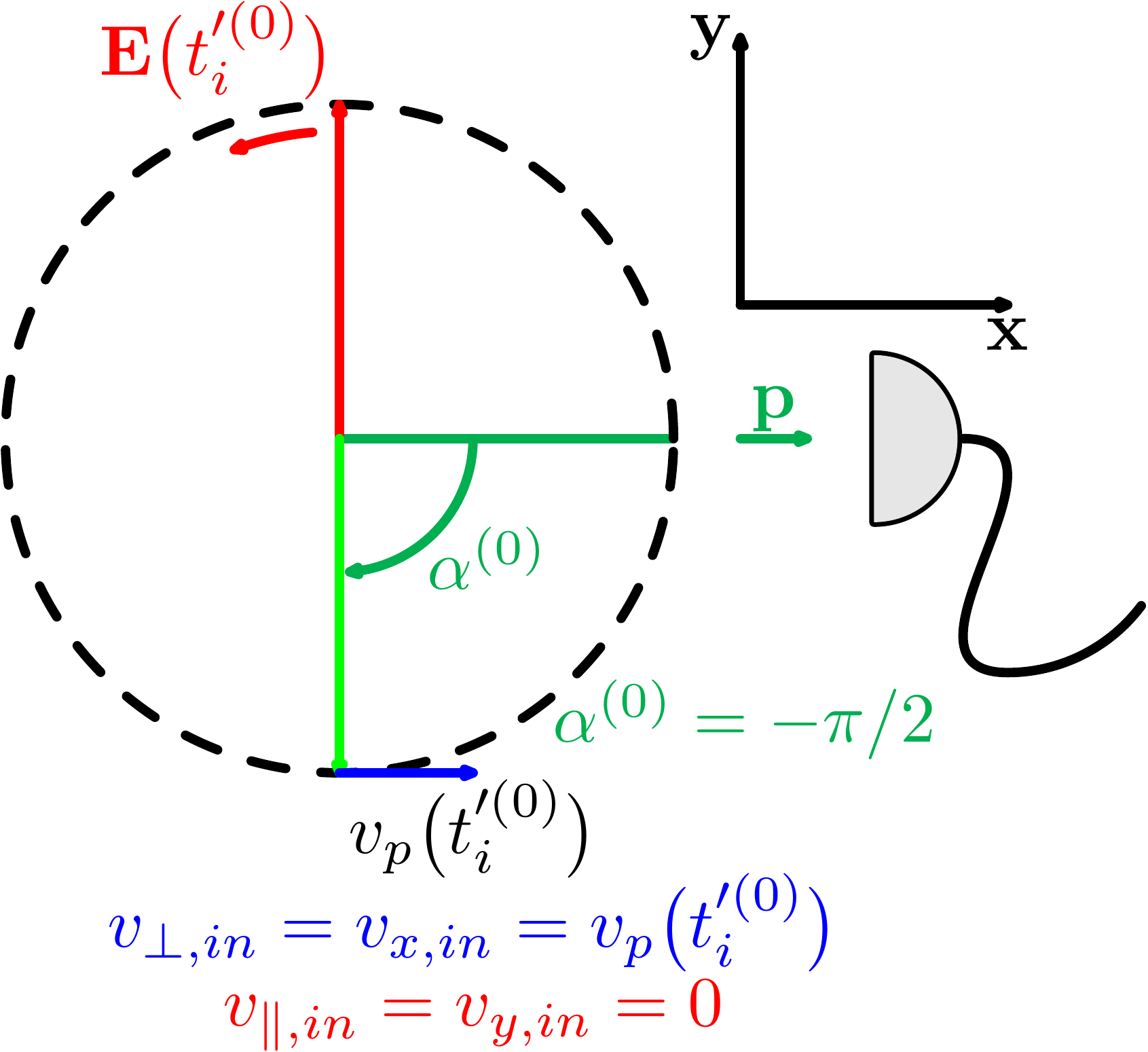} \label{fig:zrp}}\qquad
	\subfigure[]{\includegraphics[scale=0.4]{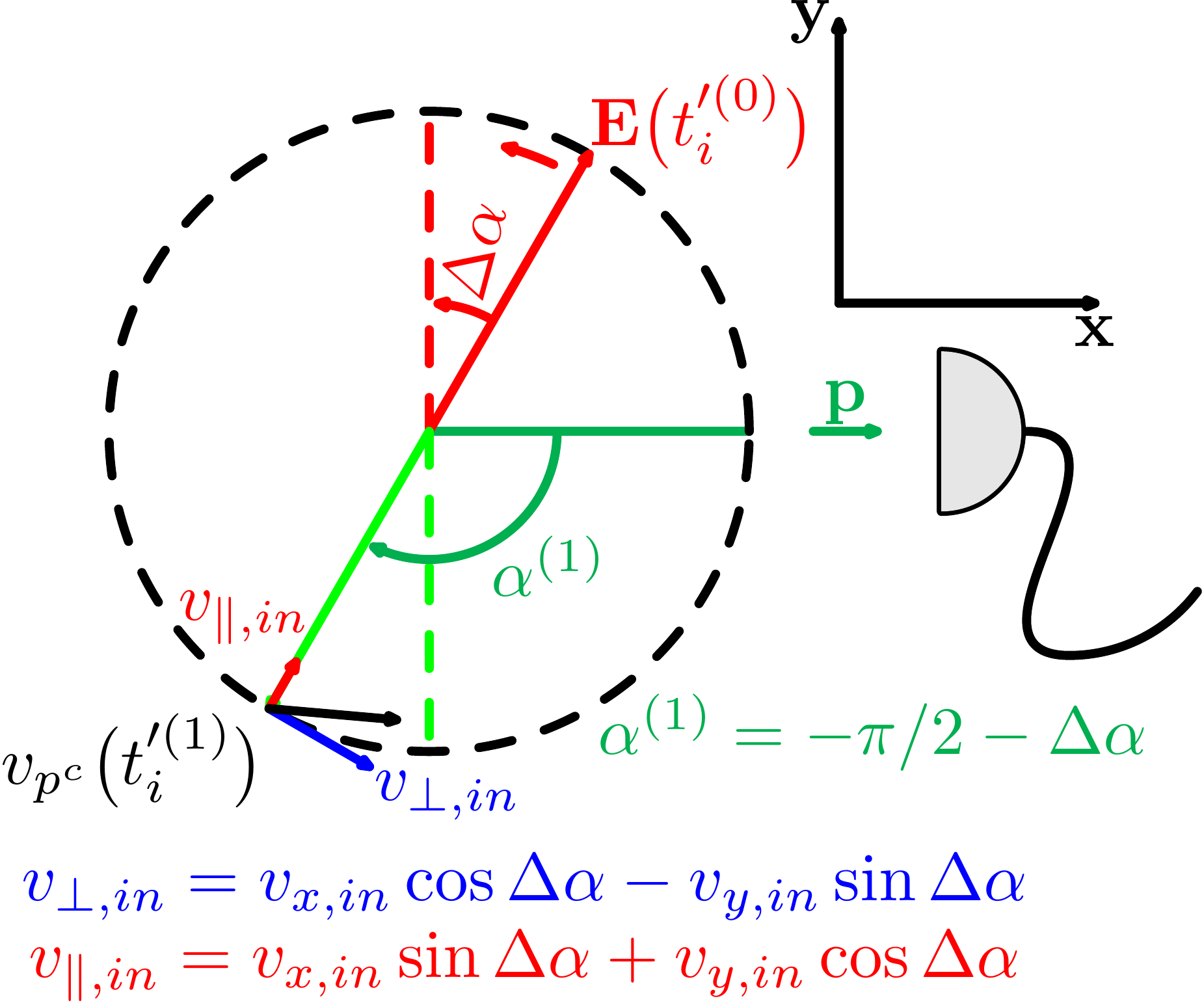}
\label{fig:lrp}}
	\caption{(Color online) Kinematics of electron tunneling through the rotating barrier. The right circularly polarized laser field $E$ creates a tunneling barrier rotating counter-clockwise. (a) Short-range potential: The electron observed at the detector placed along the $x$ axis, exits the barrier along the negative direction of the $y$ axis at angle $\alpha^{(0)} = -\pi/2$. (b) Long-range potential: The electron observed at the detector placed along the $x$ axis, exits the barrier at the angle $\alpha^{(1)} = -\pi/2 - \Delta\alpha$, $\Delta\alpha = \left|\omega\Delta t_i^{\prime(0)}\right|$, and $\Delta t_i^{\prime(0)} < 0$.} \label{fig:Physical_picture}
\end{figure}

Simple analytical calibration can be made if one neglects the electron interaction with the long-range core potential during and after ionization. For short-range potentials the mapping is illustrated in Fig.~\ref{fig:Physical_picture}. For the laser field defined as
	\begin{equation}
		\mathbf{E}(t) = E_0(-\sin(\omega t)\,\hat{\mathbf{x}} + \cos(\omega t)\,\hat{\mathbf{y}}), \label{field}
	\end{equation}
the connection between the real part of the ionization time and  the observation angle is \cite{ppt1966, barth2013}:
	\begin{equation}
		\omega t_i^{\prime(0)} = \omega\Re\left[t_s^{\prime(0)}\right]=\phi_p+2\pi(r-1), \,r \in \mathbb{N} \label{sfa_time}
	\end{equation}
The detector placed along the positive direction of the $x$ axis will detect the electron liberated at $t_i^{\prime(0)} = 0$, i.e., when the laser field $\mathbf{E}(t) = E_0\hat{\mathbf{y}}$ is pointing towards the positive direction of the $y$ axis. The electron exits the barrier in the negative direction of the $y$ axis, corresponding to the angle $-\pi/2$. The velocity at the exit, $v_y\left(t_i^{\prime(0)}\right) = 0$, $v_x\left(t_i^{\prime(0)}\right) = p_{\text{opt}} - A_0$, and $v_x\left(t_i^{\prime(0)}\right)$, tends to 0 in the tunneling limit ($\gamma \ll 1$): $v_x\left(t_i^{\prime(0)}\right) = \sqrt{2I_p}\gamma/6$. Thus, the angle between the direction of the field at the moment of ionization and the electron momentum at the detector is $\pi/2$.

How is this mapping affected when the interaction with the long-range core potential is taken into account?

\subsection{Coulomb correction to the ionization time, initial electron velocity}

Even in the tunneling limit, our analysis shows that due to the effects of the long-range potential,
the electron has nonzero velocity $(-\Delta p_y^{\text{re}})$ in the negative direction of the $y$ axis when the field is pointing in the positive $y$ direction, i.e., at $t = 0$ in our notations. This is by no means surprising and the corresponding velocity has a very simple explanation: it is required to overcome the attraction of the Coulomb potential, which the electron will experience all the way towards the detector. Had the electron been born with zero velocity the in long-range potential, it would never have reached the detector placed in the positive direction of the $x$ axis. One expects the same result within the adiabatic tunneling picture. The question is: Is the magnitude of $\Delta p_y^{\text{re}}$ consistent with the adiabatic ionization model, which would suggest that the electron was liberated slightly before $t = 0$ but with zero velocity?

To answer this question, we need to analyze the changes in the ionization time due to
the effects of the long-range potential. The corrections to ionization times associated with electron interaction with the long-range potential are given by Eqs.~\eqref{im_time} and \eqref{re_time}. The shift of the saddle point in time $\Re\left[\Delta t_s^{\prime(0)}\right]$ corresponds to the shift in the direction of the force of the electric field $-\mathbf{E}(t)$ from $-\pi/2$ to $-\pi/2 + \omega\Re\left[\Delta t_s^{\prime(0)}\right]$.

\begin{figure}
	\includegraphics[scale=0.5]{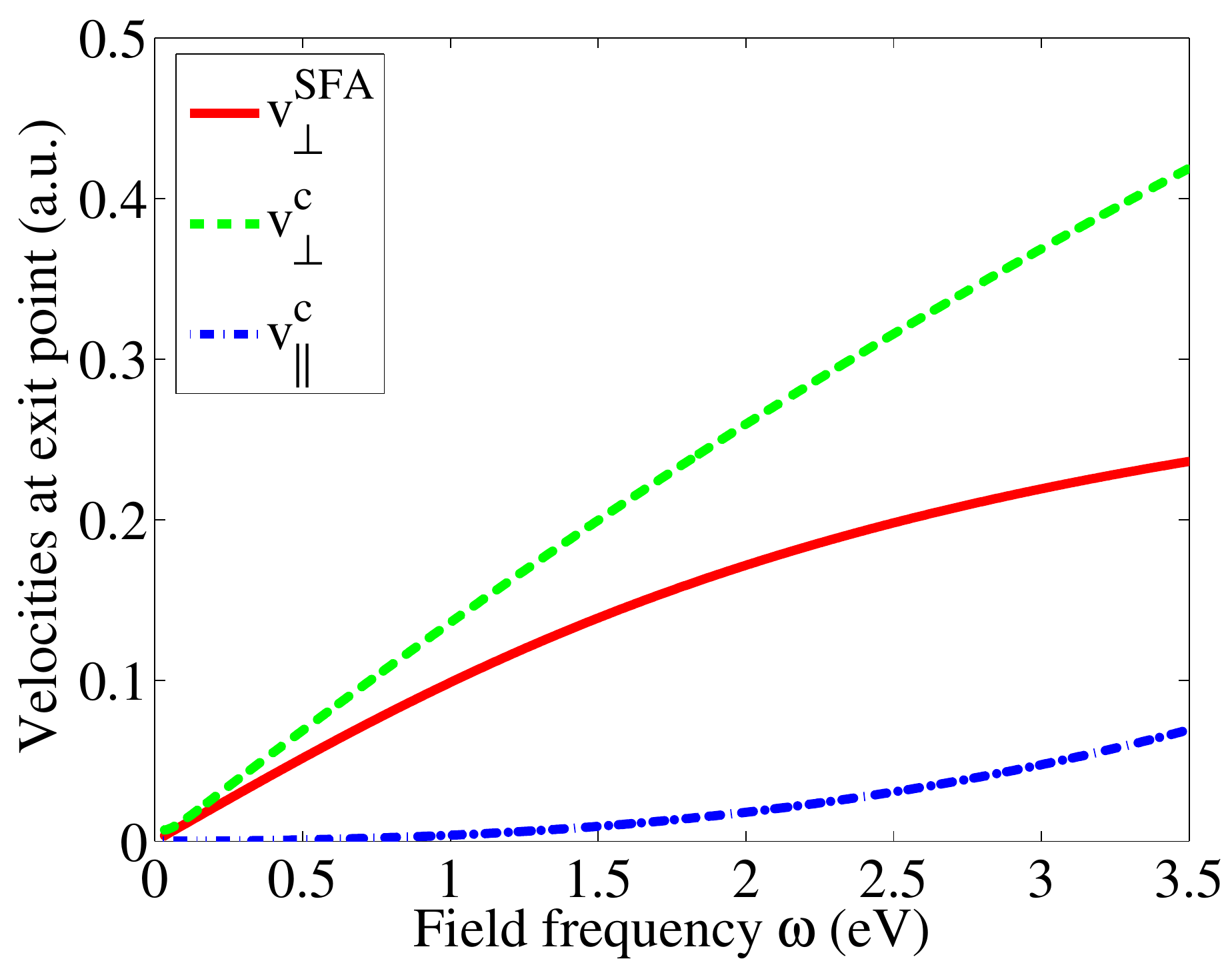}
	\caption{(Color online) Initial velocity corresponding to the center of the velocity distribution: $v_{\perp}$ [dashed (green) line] [Eq.~\eqref{coulomb exit velocityx1}] and $v_{\parallel}$ [dot-dashed (blue) line] [Eq.~\eqref{coulomb exit velocityy1}] vs frequency for $E_0 = 3 \times 10^{10}$ V/m ($E_0 = 0.06$ a.u. and $I = 2.6 \times 10^{14}$ W/cm$^{2}$) and $I_p = 14$ eV.  $v_{\perp}^{\text{SFA}}$ [solid (red) line] shows the result arising in the nonadiabatic short-range theory (the PPT theory; see \cite{ppt1966,barth2011,barth2013}) and in the length-gauge SFA.} \label{fig:initial_velocity}
\end{figure}

Let us first discuss the initial conditions for the electron continuum dynamics in the tunneling limit $\gamma \ll 1$. In this limit, the electron moves in static electric field $[\mathbf{E}(t) = E_0\,\hat{\mathbf{y}}]$ and the momentum shift is accumulated along the electron trajectory,
	\begin{equation}
		{y}^{\text{tun}}(t)  = -\left[\frac{I_p}{E_0} + \frac{1}{2}E_0\left(t - t_i^{\prime(0)}\right)^2\right]\hat{\mathbf{y}},
	\end{equation}
where ${y}^{\text{tun}}\left(t_i^{\prime(0)}\right) = -I_p/E_0$ is the coordinate of the exit point in the tunneling limit. Taking into account that $U = -Q/(-y)$, $\nabla U = -Q\hat{\mathbf{y}}/y^2$ and substituting this trajectory into the expression for $\Delta\mathbf{p}$, Eq.~\eqref{Dp_realfin}, we obtain
	\begin{align}
		\Re[\Delta p_y] &= -Q\int_{t_i^{\prime(0)}}^{T}\frac{d\tau}{(y^{\text{tun}})^2} = -\frac{0.78\sqrt{2}}{I_p^{3/2}}QE_0. \label{dpytun}		
	\end{align}
It is easy to see that Eq.~\eqref{re_time} in the tunneling limit yields $\Re\left[\Delta t_s^{\prime(0)}\right] = -\Delta p_y^{\text{re}}/E_0$, thus we obtain from Eq.~\eqref{dpytun}:
	\begin{align}
		\Delta t_i^{\prime(0)} = -\frac{0.78Q\sqrt{2}}{I_{p}^{3/2}} &\approx -I_p^{3/2}. \label{dti}
	\end{align}
From Eq.~\eqref{dti}, we find that the correction to the ionization time $\Re\left[\Delta t_s^{\prime(0)}\right]$ is negative, the electron is born before $\mathbf{E}(t)$ points down, and the Coulomb corrected angle $-\pi/2 + \omega\Re\left[\Delta t_s^{\prime(0)}\right]$ has a negative value. At this (earlier) ionization time the electron velocity is lower than at $t_i^{\prime(0)}$, and in the tunneling limit:
	\begin{align}
		v_x &= p_{\text{opt}} - A_0\cos\left(\omega t_i^{\prime(0)} + \omega\Re[\Delta t_s^{\prime(0)}]\right) - \Delta p_x \approx p_{\text{opt}} - A_0 + \mathcal{O}(G_C^2) \approx \gamma\sqrt{2I_p}/6 + \mathcal{O}(G_C^2),\\
		v_y &= -\Delta p_y - A_0\sin\left(\omega t_i^{\prime(0)} + \omega\Re[\Delta t_s^{\prime(0)}]\right) = -\Delta p_y - A_0\omega\Re[\Delta t_s^{\prime(0)}] \approx 0 + \mathcal{O}(G_C^2).
	\end{align}
Thus, in the tunneling limit $\gamma \to 0$, the electron velocity indeed tends to 0 at the exit from the barrier. The effect of the Coulomb potential is reduced to the modification of the angle between the direction of the laser field at the moment of exit $\mathbf{E}\left(t_i^{\prime(0)}\right)$ and the direction of the final electron momentum $\mathbf{p}$, registered at the detector. For short-range potentials this angle is $\pi/2$, and for long-range potentials this angle is larger; in the tunneling limit it is $\pi/2 + \omega I_p^{-3/2}$, (see Fig.~\ref{fig:Physical_picture}).

However, most of the experiments are currently performed in the regime of nonadiabatic ionization, when the Keldysh parameter $\gamma$ is not that small. In this regime the exit velocities (with $t_i^{\prime(0)} = 0$),
	\begin{align}
		v_x &= p_{\text{opt}} - A_0\cos\left(\omega\Re[\Delta t_s^{\prime(0)}]\right) - \Delta p_x, \label{coulomb_exit_velocityx}\\
		v_y &= -\Delta p_y - A_0\sin\left(\omega\Re[\Delta t_s^{\prime(0)}]\right), \label{coulomb_exit_velocityy}
	\end{align}
become significant already for small $\gamma$. The longitudinal electron velocity $v_{\parallel}$ along the direction of the field and the transverse electron velocity $v_{\perp}$ orthogonal to the field are also non-zero (Fig.~\ref{fig:initial_velocity}). The longitudinal and transverse velocities are obtained from Eqs.~\eqref{coulomb_exit_velocityx} and \eqref{coulomb_exit_velocityy} ($\Delta \alpha=\left|\omega \Delta t_i^{\prime(0)}\right|$):
	\begin{align}
		v_{\perp} &= v_x\cos\left(\Delta \alpha\right) - v_y\sin\left(\Delta \alpha\right), \label{coulomb exit velocityx1}\\
		v_{\parallel} &= v_x\sin\left(\Delta \alpha\right) + v_y\cos\left(\Delta \alpha\right). \label{coulomb exit velocityy1}
	\end{align}

Ignoring the non-zero initial velocity of the electron will generally lead to errors in the two-step reconstruction of time delays in the angular streaking method. In the next section we illustrate the degree of uncertainty that can arise in reconstructing the time from the attoclock  measurement using examples of Ar and He atoms.

\subsection{Calibration of the attoclock}

The attoclock observable is the angular offset. This angular offset either can appear due to electron interaction with the core potential $\Delta\alpha$, as described above, or can be associated with other delays, e.g., delays accumulated due to nontrivial tunneling, polarization, or excitation dynamics, $\Delta\alpha^U$ (the superscript $U$ stands for ``unknown," since the respective $\Delta\alpha^U$ is associated with the dynamics that we may not know).
Since the attoclock can only measure the total offset $\Delta\alpha^{T} = \Delta\alpha^{U} + \Delta\alpha$, to get access to the unknown (e.g., tunneling) times one has to calculate the offset $\Delta\alpha$ and subtract it from the measurable offset $\Delta\alpha^{T}$. The uncertainty in the calculation of $\Delta \alpha$ will lead to the corresponding uncertainty in reconstructing, say, the tunneling time.

\begin{figure}
	\centering
	\begin{tabular}{cc}
		\subfigure[]{\includegraphics[scale=0.4]{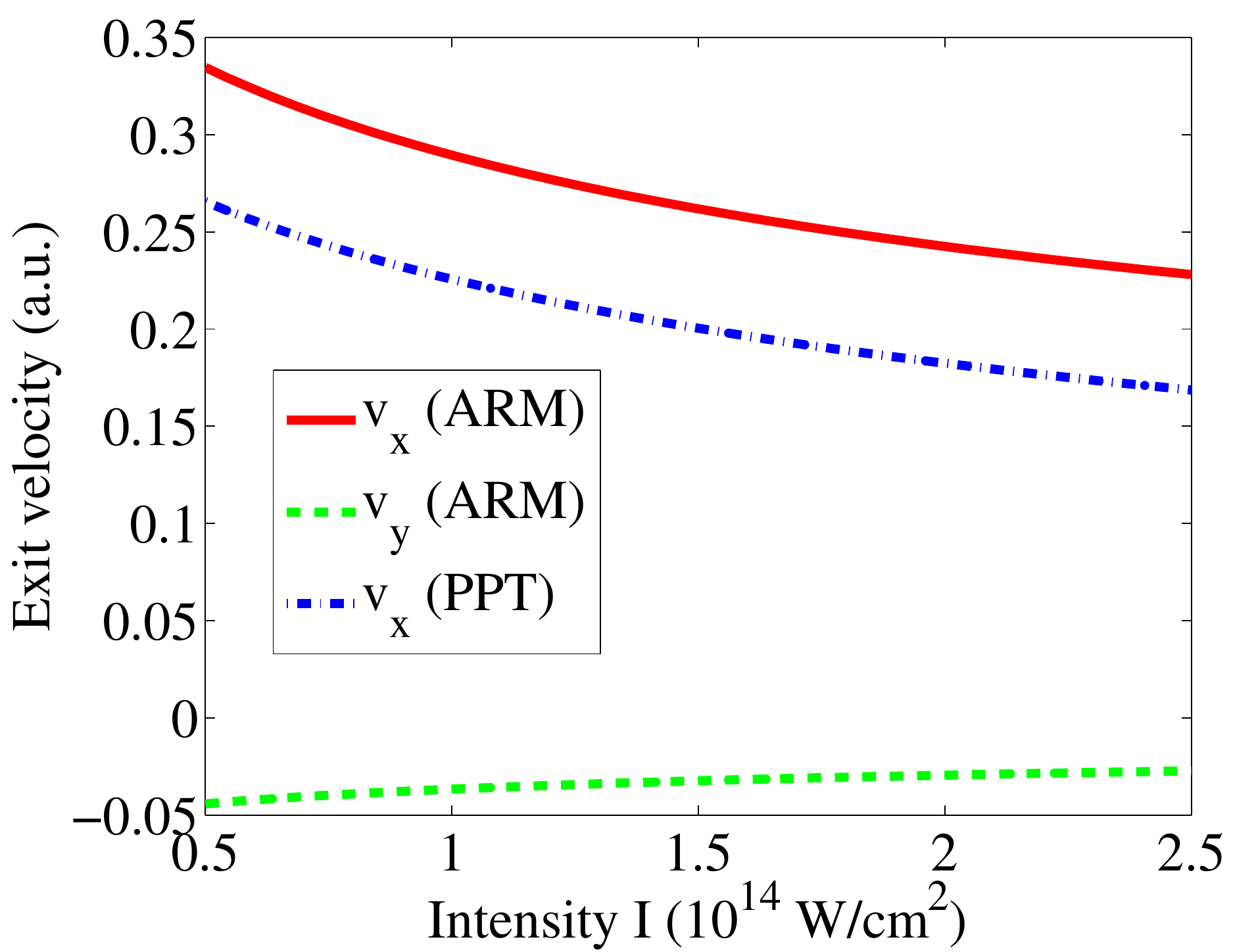}\label{fig:initial_vel_Ar}}\hspace{0.5cm} &
		\subfigure[]{\includegraphics[scale=0.4]{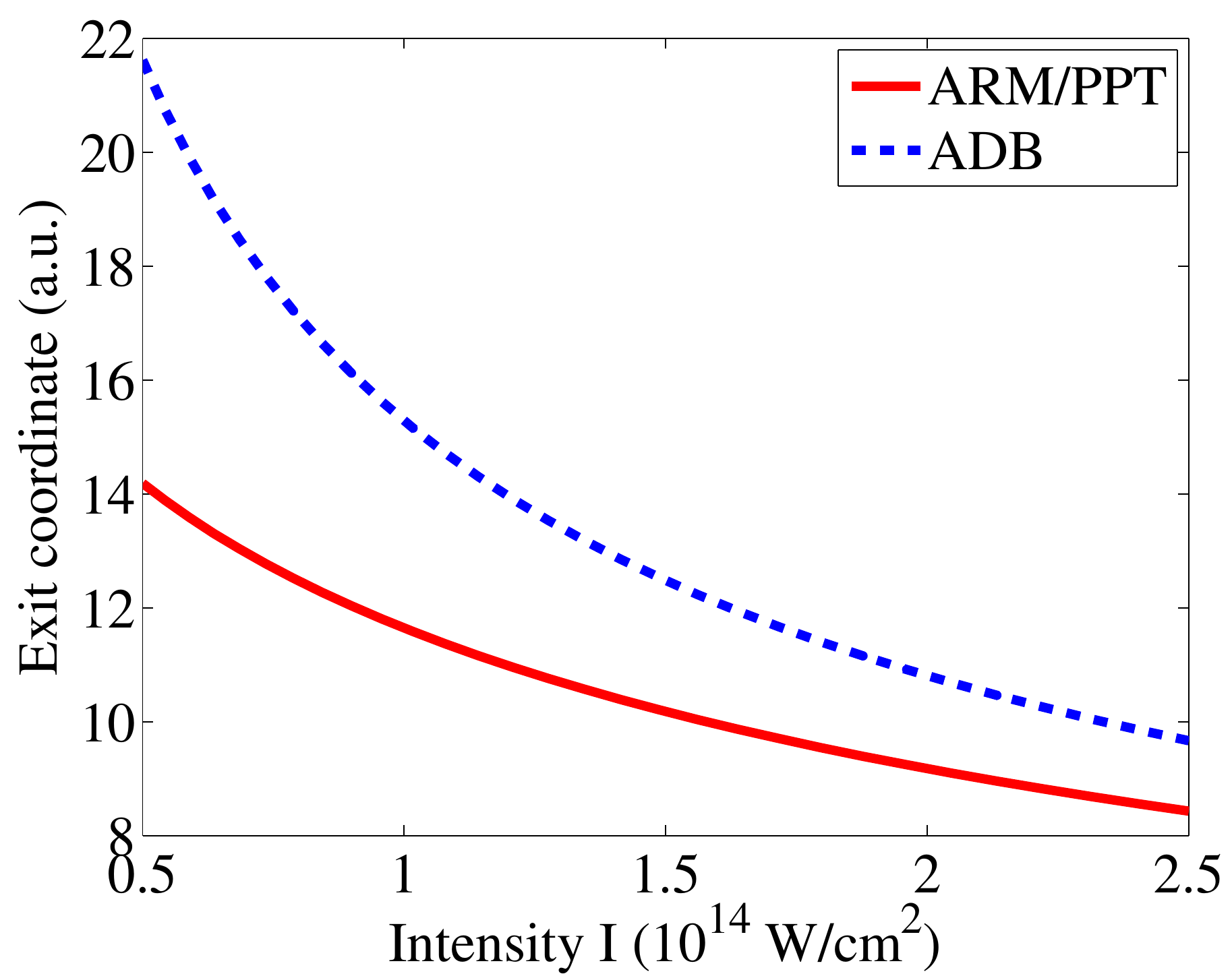}\label{fig:initial_coord_Ar}}\\
		\subfigure[]{\includegraphics[scale=0.4]{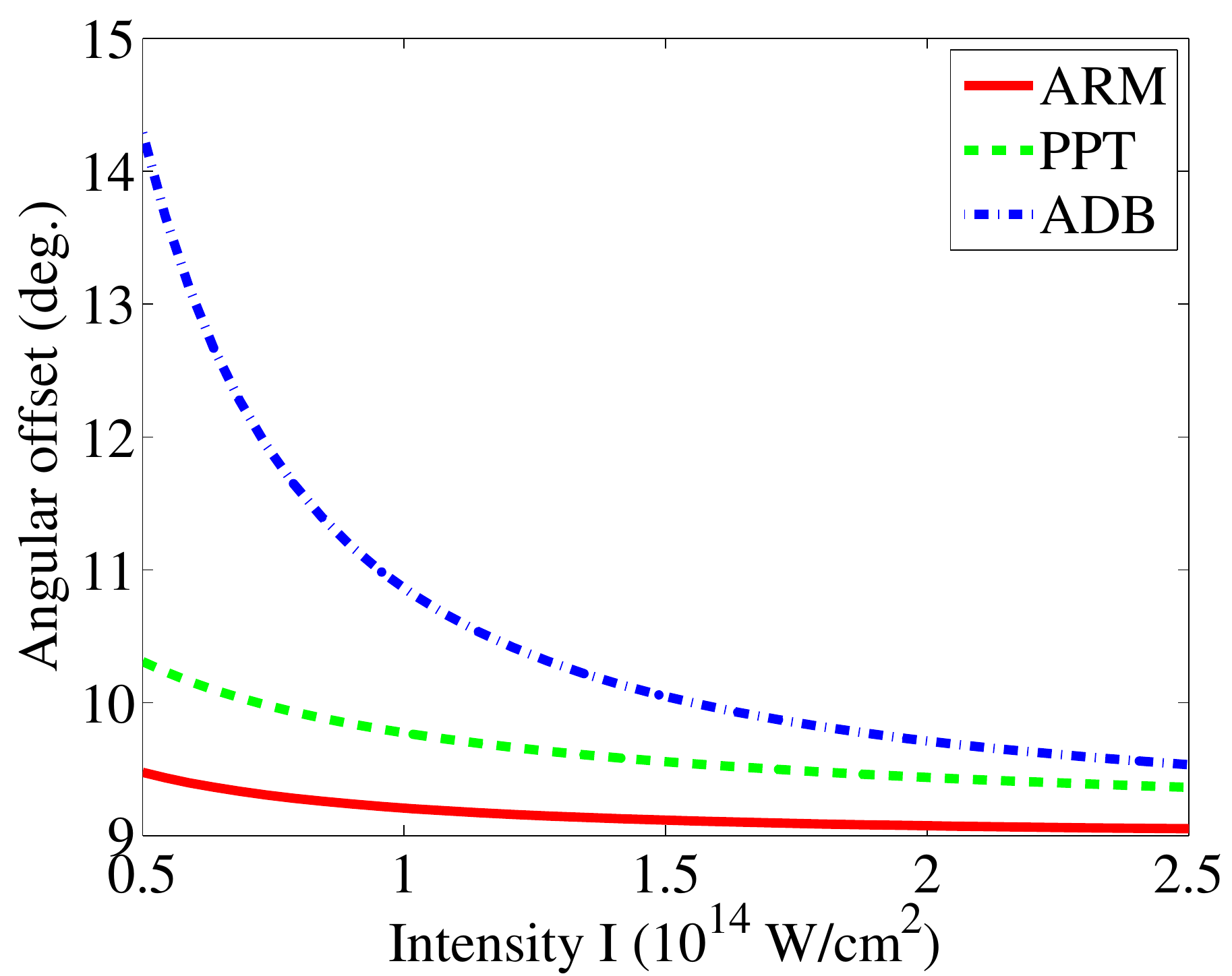}\label{fig:angular_offset_Ar}}\hspace{0.5cm} &
		\subfigure[]{\includegraphics[scale=0.4]{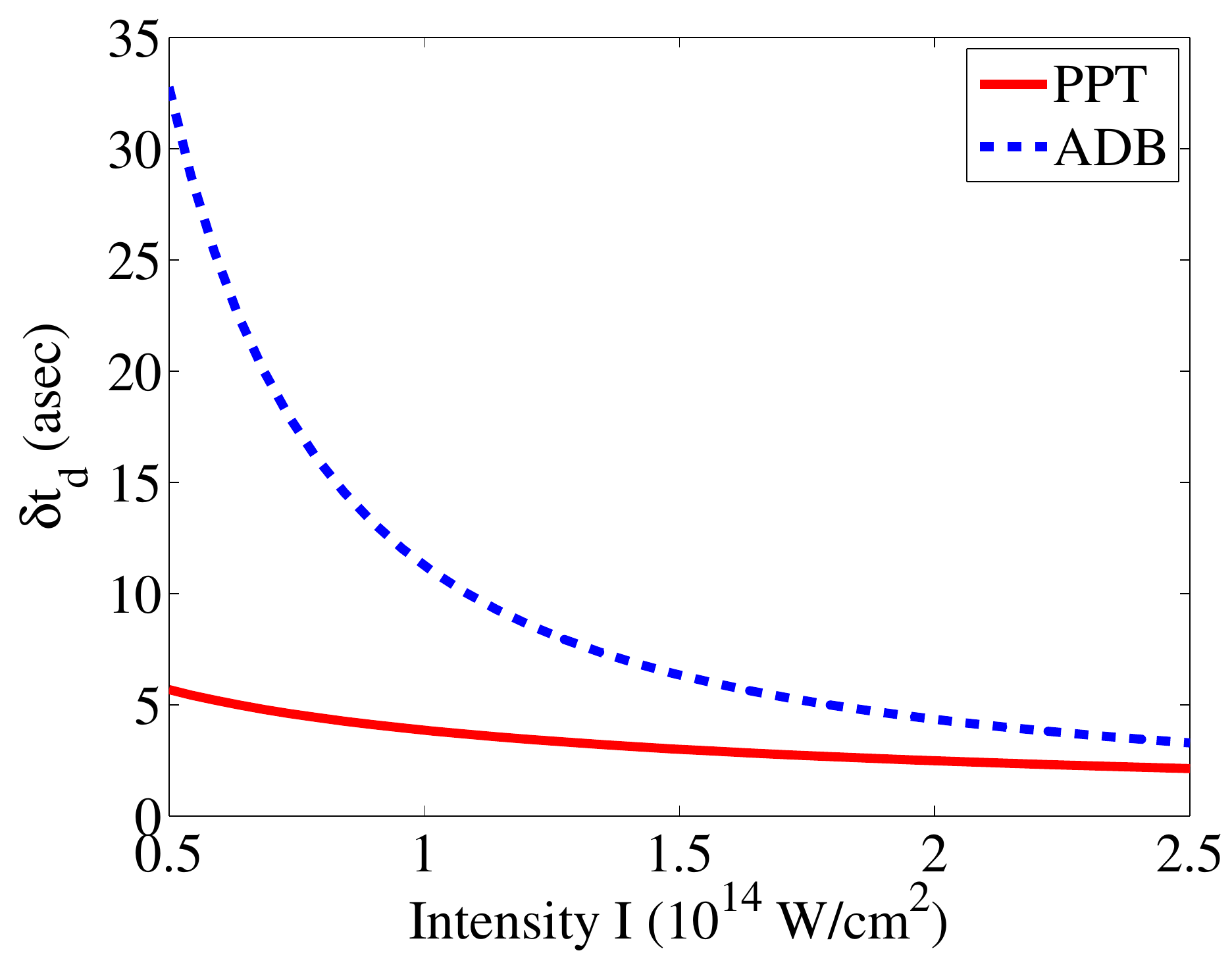}\label{fig:angular_offset_diff_Ar}}
	\end{tabular}
	\caption{(Color online) Calibration of the attoclock for an Ar atom with $I_p = 15.7$ eV.
  (a) Initial velocities  $v_x$ [solid (red) curve] and $v_y$ [dashed (green) curve] resulting from the A$R$M theory and $v_x$ [dot-dashed (blue) curve] from the nonadiabatic short-range theories \cite{ppt1966, barth2011, barth2013} for the geometry specified in Fig.~\ref{fig:Physical_picture}.
  (b) Initial coordinate (exit point) in the A$R$M and the PPT \cite{ppt1966,barth2011,barth2013} theories [solid (red) curve], and $I_p/E_0$ in the adiabatic theory [dashed (blue) curve].
  (c) Angular offset $\Delta\alpha$ corresponding to the A$R$M [solid (red) curve], nonadiabatic short-range [dashed (green) curve] and adiabatic [dot-dashed (blue) curve] theories.
  (d) Uncertainty in the calibration of time in the attoclock corresponding to (i) the nonadiabatic two-step model [solid (red) curve] and (ii) adiabatic two-step model [dashed (blue) curve].}
\end{figure}

In this section we consider the angular offset $\Delta\alpha$ and analyze the associated uncertainties in the time reconstruction for three models.
\begin{enumerate}[i.]
	\item {\it The two-step adiabatic model}. This model assumes that the peak of the photoelectron distribution corresponds to the electron trajectory with specific initial conditions, namely, the initial coordinate defined according to the quasistatic tunneling picture for short-range potentials, or in the limit of a sufficiently thick barrier ($4E_0\ll I_p^2$): $x_{e}^{\text{qs}}\left(t_i^{\prime(0)}\right) = 0$, $y_{e}^{\text{qs}}\left(t_i^{\prime(0)}\right) = -I_p/E_0$. The initial electron velocity is 0 (both transversal and longitudinal): $v_x^{\text{qs}}\left(t_i^{\prime(0)}\right) = 0$, $v_y^{\text{qs}}\left(t_i^{\prime(0)}\right) = 0$.
	\item {\it The two-step nonadiabatic model}. The peak of the photoelectron distribution corresponds to the electron trajectory. The initial coordinate is defined according to the PPT theory $y_e^{\prime(0)} = \int_{t_s^{\prime(0)}}^{\Re\left[t_s^{\prime(0)}\right]}d\zeta\,\left[{p}_{\text{opt}} + {A}_y(\zeta)\right]$ [see also Eq.~\eqref{opt_coord_exit}]. The initial electron velocity is nonzero in the direction orthogonal to the field polarization at the time of exit: $v_x \left(t_i^{\prime(0)}\right) = p_{\text{opt}} - A_0$, $v_y\left(t_i^{\prime(0)}\right) = 0$ [see Eqs.~\eqref{optimal_momentum} and \eqref{optimal_momentum1} for the definition of $p_{\text{opt}}$]. This nonzero velocity reflects the presence of a ``cross-wind": the effect of the second component of the circuarly polarized field. Note that, both orthogonal components of the circular field are always non-zero in sub-barrier region, when electron trajectory evolves in complex time.
	\item {\it A$R$M model}. The A$R$M model is a consistent quantum approach which does not require the knowledge of the ``initial conditions"  to calculate the offset angle. However, since the A$R$M method naturally incorporates the concept of trajectories, the initial conditions can be obtained within the A$R$M model, as discussed in the previous sub-section. Both $v_x\left(t_i^{\prime(0)}\right)$ and $v_y\left(t_i^{\prime(0)}\right)$ are nonzero due to the nonadiabatic Coulomb effects [see Eqs.~\eqref{coulomb_exit_velocityx} and \eqref{coulomb_exit_velocityy}].
\end{enumerate}

To ensure that all three models use the same level of approximation for the electron continuum dynamics, in two-step models we propagate the trajectories from the point of exit to the detector using the EVA instead of solving Newton's equations exactly. Formally, this means that the classical equation for Coulomb pluse laser field (used in the two-step model),
	\begin{equation}
		\frac{d\mathbf{r}}{dt} = \mathbf{v}(t),\quad \frac{d\mathbf{v}}{dt} = -\mathbf{E}(t) - \frac{Q}{r^3(t)}\mathbf{r}(t)
	\end{equation}
$\mathbf{r}(t) = \left(x(t), y(t)\right)$, is solved iteratively. The zeroth-order trajectory (neglecting the Coulomb term) is used in the argument of the Coulomb potential. For the two-step adiabatic model, we obtain
	\begin{align}
		v_{x}^{\text{ADB}} = A_0 + \Delta p_{x}^{\text{ADB}},\quad v_{y}^{\text{ADB}} = \Delta p_{y}^{\text{ADB}},\quad \Delta\alpha^{\text{ADB}} = \operatorname{\tan^{-1}}\left(\frac{v_{y}^{\text{ADB}}}{v_{x}^{\text{ADB}}}\right),
	\end{align}
where $\Delta p_{x}^{\text{ADB}}$ and $\Delta p_{y}^{\text{ADB}}$ are defined as ($\phi = \omega t$, $\phi_{T} = \omega T$, $T \to \infty$)
	\begin{gather}
		\Delta p_{x}^{\text{ADB}} = -\frac{Q\omega}{A_0^2}\int_{0}^{\phi_{T}}d\phi\,\frac{x^{\text{ADB}}(\phi)}{\left[\left(x^{\text{ADB}}(\phi)\right)^2 + \left(y^{\text{ADB}}(\phi)\right)^2\right]^{\frac{3}{2}}},\\
		\Delta p_{y}^{\text{ADB}} = -\frac{Q\omega}{A_0^2}\int_{0}^{\phi_{T}}d\phi\,\frac{y^{\text{ADB}}(\phi)}{\left[\left(x^{\text{ADB}}(\phi)\right)^2 + \left(y^{\text{ADB}}(\phi)\right)^2\right]^{\frac{3}{2}}},\\
		x^{\text{ADB}}(\phi) = -\sin\phi + \phi,\\
		y^{\text{ADB}}(\phi) = \cos\phi - 1 - \gamma^2/2.
	\end{gather}
For the two-step nonadiabatic model, we obtain
	\begin{align}
		v_{x}^{\text{PPT}} = p_{\text{opt}} + \Delta p_{x}^{\text{PPT}},\quad v_{y}^{\text{PPT}} = \Delta p_{y}^{\text{PPT}},\quad \Delta\alpha^{\text{PPT}} = \operatorname{\tan^{-1}}\left(\frac{v_{y}^{\text{PPT}}}{v_{x}^{\text{PPT}}}\right),
	\end{align}
where $\Delta p_{x}^{\text{PPT}}$ and $\Delta p_{y}^{\text{PPT}}$ are defined as
	\begin{gather}
		\Delta p_{x}^{\text{PPT}} = -\frac{Q\omega}{A_0^2}\int_{0}^{\phi_{T}}d\phi\,\frac{x^{\text{PPT}}(\phi)}{\left[\left(x^{\text{PPT}}(\phi)\right)^2 + \left(y^{\text{PPT}}(\phi)\right)^2\right]^{\frac{3}{2}}},\\
		\Delta p_{y}^{\text{PPT}} = -\frac{Q\omega}{A_0^2}\int_{0}^{\phi_{T}}d\phi\,\frac{y^{\text{PPT}}(\phi)}{\left[\left(x^{\text{PPT}}(\phi)\right)^2 + \left(y^{\text{PPT}}(\phi)\right)^2\right]^{\frac{3}{2}}},\\
		x^{\text{PPT}}(\phi) = -\sin\phi + \frac{p_{\text{opt}}}{A_0}\phi,\\
		y^{\text{PPT}}(\phi) = \cos\phi - \eta(p_{\text{opt}}).
	\end{gather}
and $\eta(p_{\text{opt}})$ is given by Eq.~\eqref{eta}. Note that $x^{\text{ADB}}(\phi)$, $y^{\text{ADB}}(\phi)$ and $x^{\text{PPT}}(\phi)$, $y^{\text{PPT}}(\phi)$ are the respective trajectories in units of $E_0/\omega^2$. While this approximation can slightly affect the absolute values of the offset angles $\Delta\alpha$, the error is essentially identical for all three models. Thus, the time uncertainty, determined by the relative offset given by the two-step models with respect to the A$R$M method, is virtually unaffected.

\begin{figure}
	\centering
	\begin{tabular}{cc}
		\subfigure[]{\includegraphics[scale=0.4]{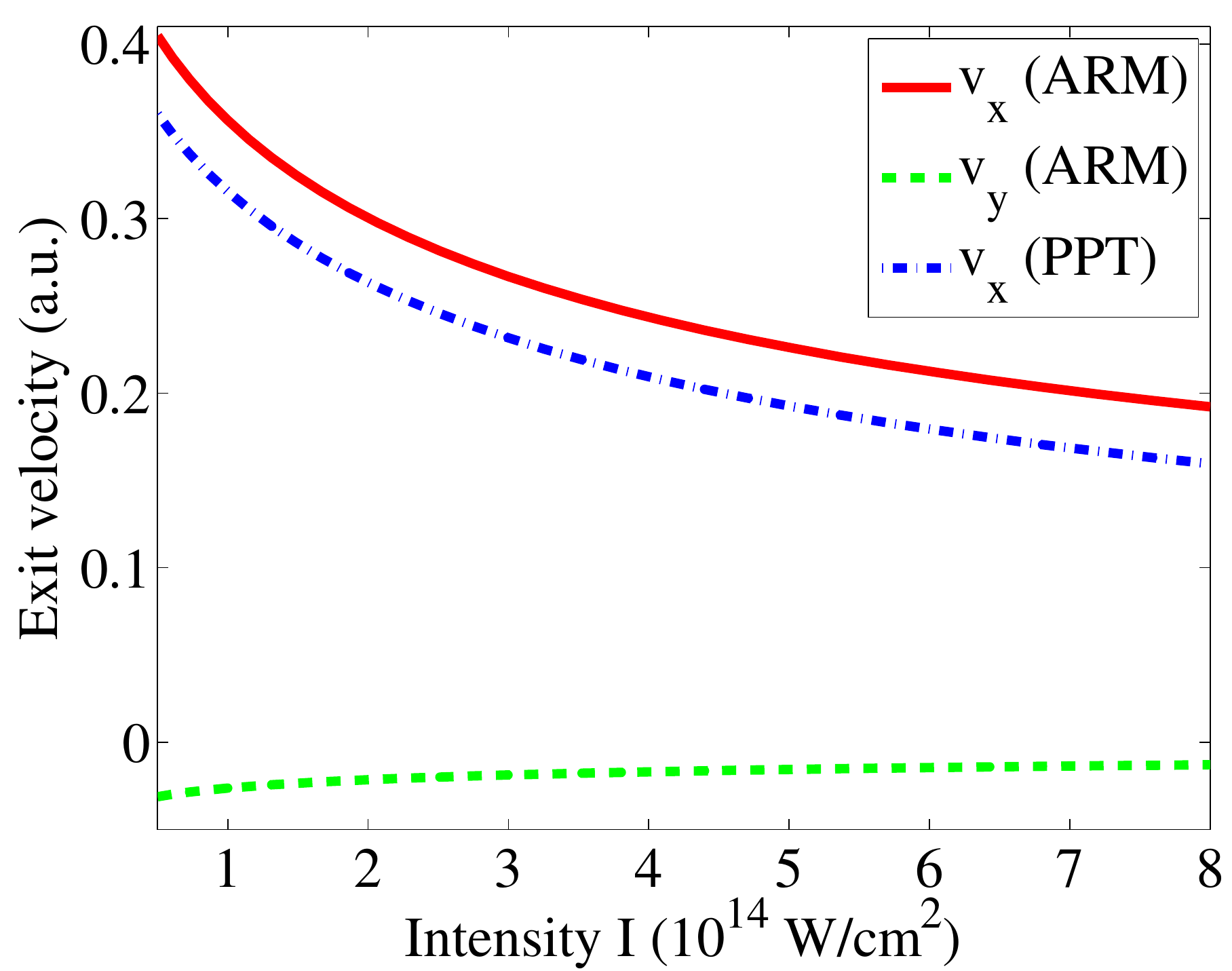}}\hspace{0.5cm} &
		\subfigure[]{\includegraphics[scale=0.4]{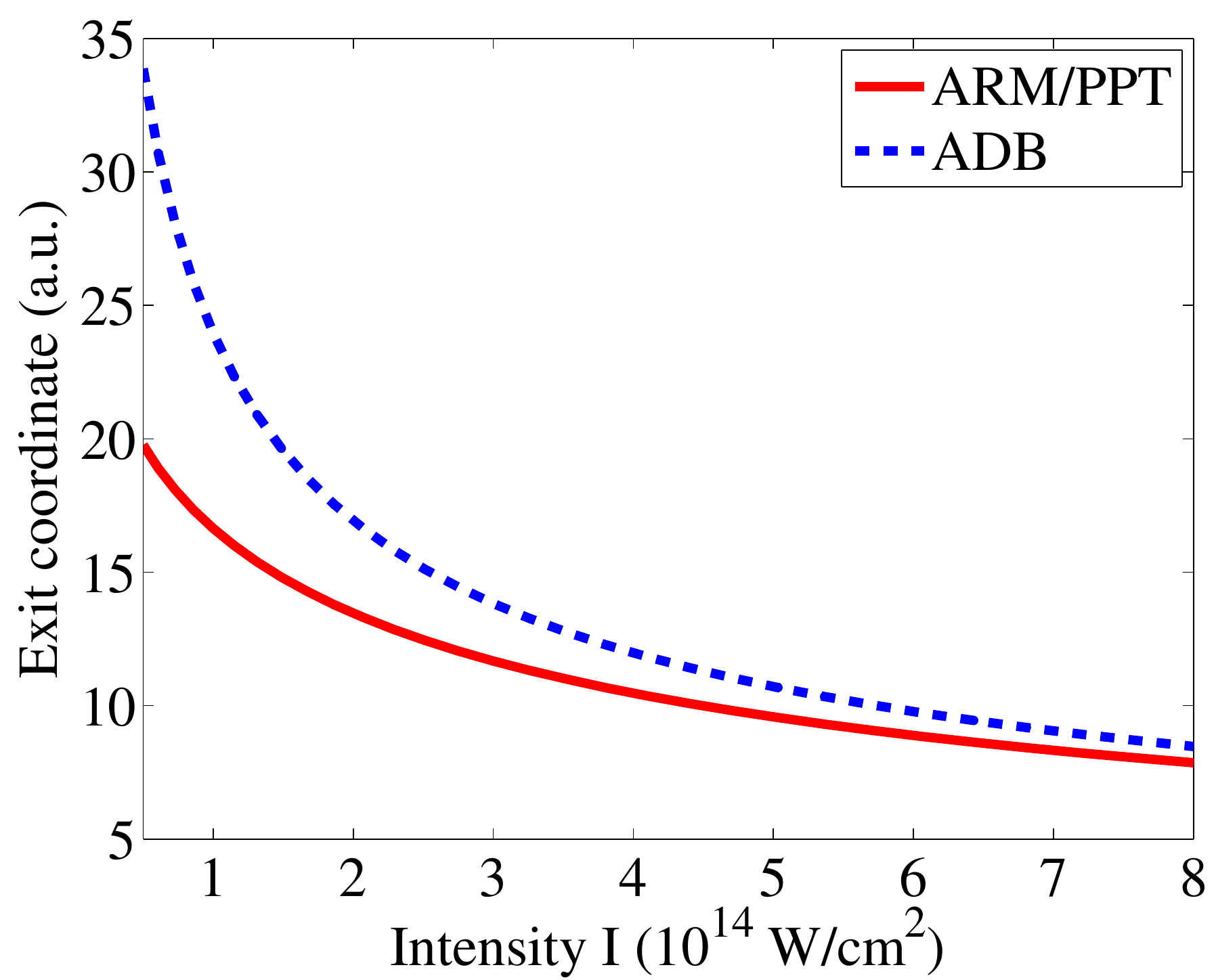}} \\
		\subfigure[]{\includegraphics[scale=0.4]{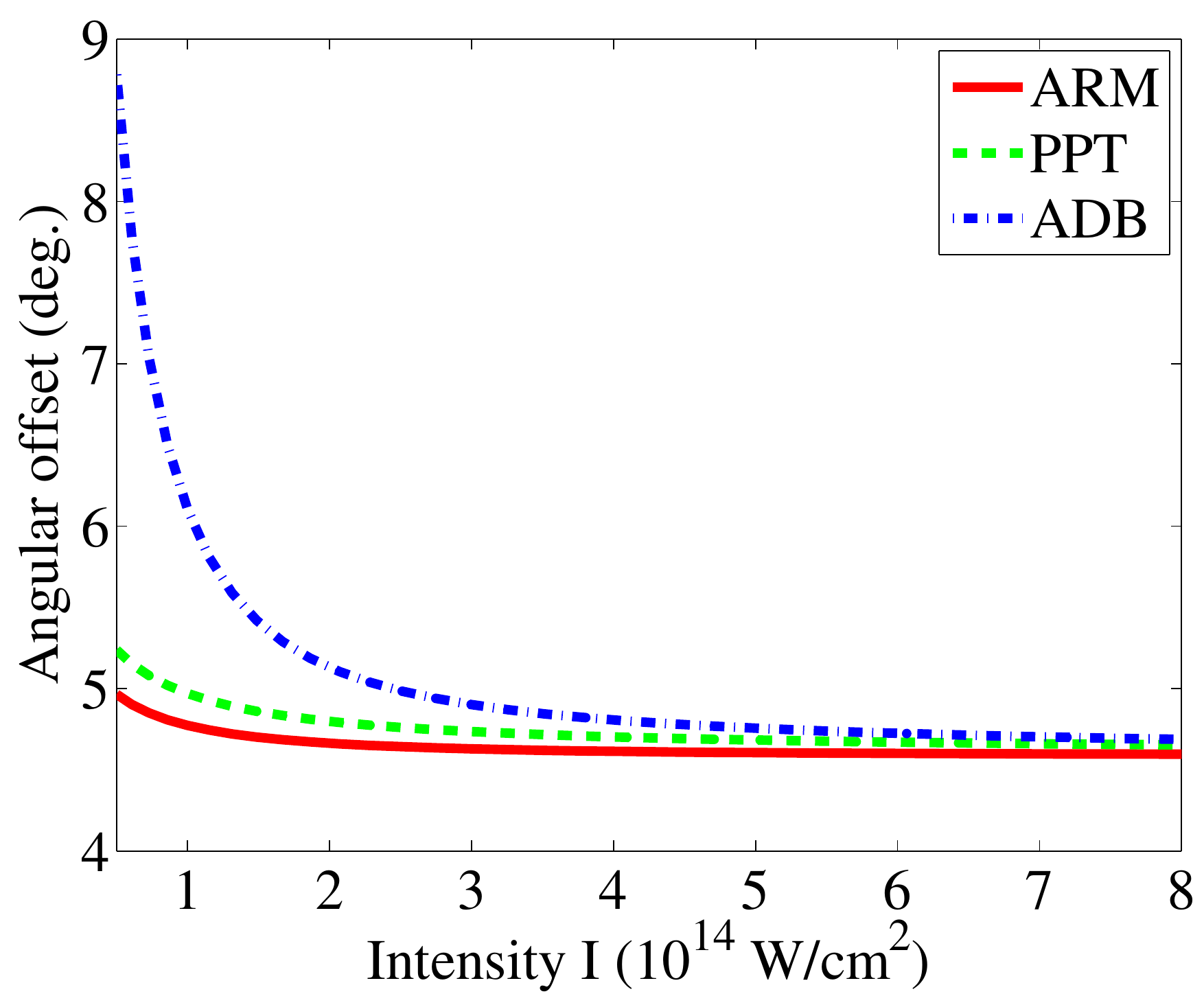}}\hspace{0.5cm} &
		\subfigure[]{\includegraphics[scale=0.4]{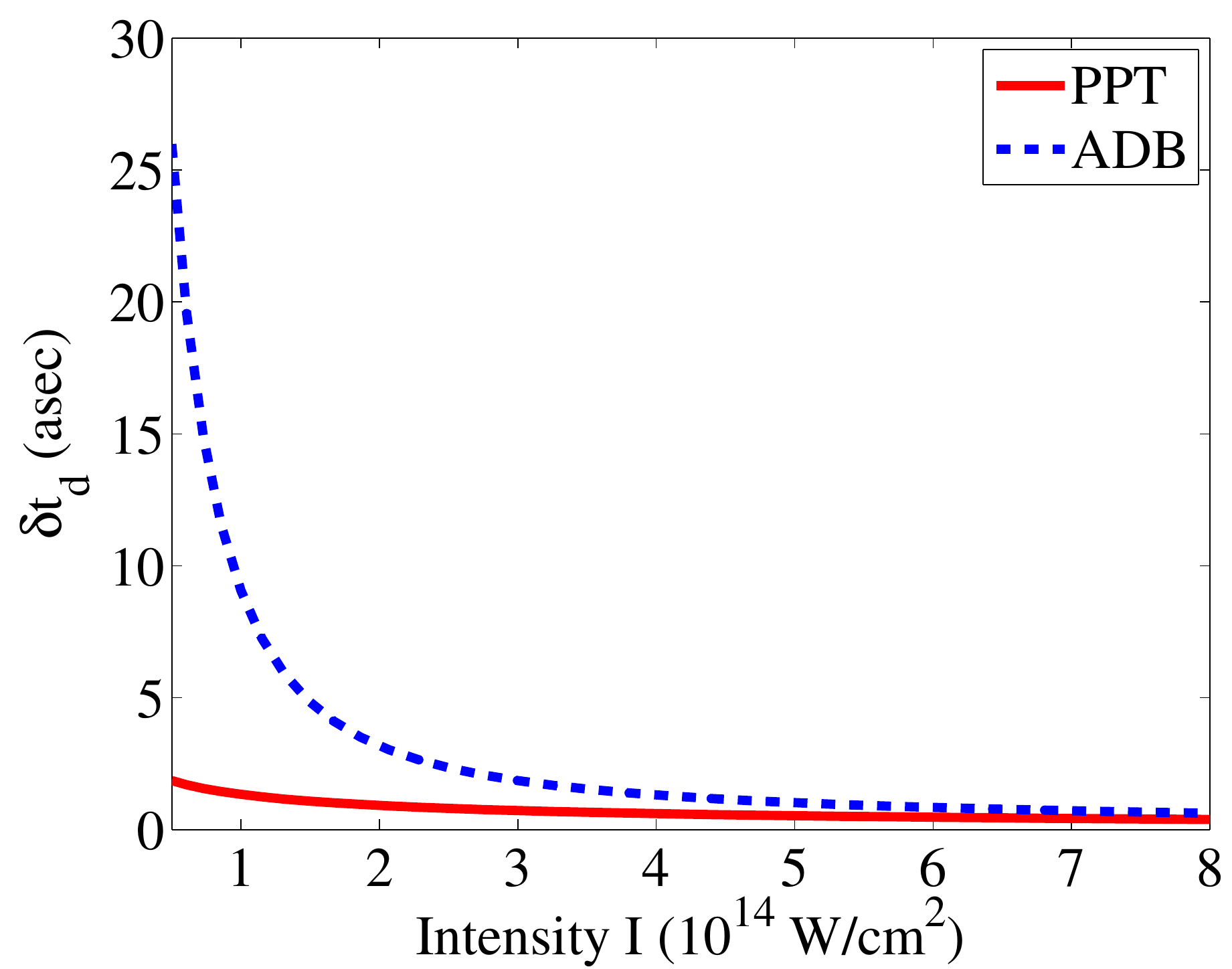}}
	\end{tabular}
	\caption{(Color online) Calibration of the attoclock for a He atom with $I_p = 24.6$ eV.
  (a) Initial velocities  $v_x$ [solid (red) curve] and $v_y$ [dashed (green) curve] resulting from the A$R$M theory and $v_x$ [dot-dashed (blue) curve] from the nonadiabatic short-range theories \cite{ppt1966, barth2011, barth2013} for the geometry specified in Fig.~\ref{fig:Physical_picture}. 
  (b) Initial coordinate (exit point) in the A$R$M and the PPT \cite{ppt1966, barth2011, barth2013} theories [solid (red) curve], and $I_p/E_0$ in the adiabatic theory [dashed (blue) curve].
  (c) Angular offset $\Delta\alpha$ corresponding to the A$R$M [solid (red) curve], nonadiabatic short-range [dashed (green) curve] and adiabatic [dot-dashed (blue) curve] theories.
  (d) Uncertainty in time reconstruction associated with the nonadiabatic [solid (red) curve] two-step model and the adiabatic two-step model [dashed (blue) curve].} \label{fig:comp_He}
\end{figure}

Figure~\ref{fig:angular_offset_Ar} shows the angular offsets for Ar atoms for all three models. The discrepancy between the models increases with the decrease in the laser intensity, reaching $\delta \alpha\approx4.5^{o}$ relative offset between the adiabatic model and the A$R$M model (for $E_0= 0.0267$ a.u., corresponding to $0.5 \times 10^{14}$ W/cm$^2$). The discrepancy is due to the different initial conditions in these models. We stress that the A$R$M theory does not require knowledge of the initial conditions to obtain $\Delta \alpha$, because it does not need to split the entire quantum process into two steps. However, the initial conditions can be obtained from the A$R$M theory, if needed.

Figure~\ref{fig:initial_vel_Ar} compares the initial velocities resulting from the A$R$M and the nonadiabatic short-range theories \cite{ppt1966, barth2011, barth2013} for the geometry specified in Fig.~\ref{fig:Physical_picture}. For the adiabatic model both $v_x$ and $v_y$ are 0 [not shown in Fig.~\ref{fig:initial_vel_Ar}]. The difference in the initial coordinates in the nonadiabatic theory for short-range potentials and the adiabatic model is shown in Fig.~\ref{fig:initial_coord_Ar}. The initial coordinate in the A$R$M model is essentially the same as in the nonadiabatic short-range theory, since the respective Coulomb correction is an order higher than the first-order Coulomb effects considered in the current implementation of the A$R$M method. The difference in the offset angle $\delta\alpha$ maps into uncertainty in the delay time: $\delta t_d = \delta\alpha/\omega$ [Fig.~\ref{fig:angular_offset_diff_Ar}]. The uncertainty in the reconstruction of the time delay becomes less significant at higher intensities and ranges from 30 as for low fields to 3 as near the barrier suppression intensity [Fig.~\ref{fig:angular_offset_diff_Ar}]. The uncertainty $\delta t_d$ strongly decreases if nonadiabatic initial conditions are used in the two-step model, ranging from 5 as for low intensities to 2 as for high intensities.

Qualitatively we find the same picture for He atoms (Fig.~\ref{fig:comp_He}), however, quantitatively the discrepancy between the different models is smaller and the time uncertainty is almost negligible for the highest intensities. For He atoms, using nonadiabatic initial conditions in the two-step model reduces the uncertainty to 1.5 as and even less for higher intensities.

%

\subsection{Coulomb correction to the electron ``tunneling angle"}

The complex tunneling angle characterizes the direction of the electron velocity at the complex ionization time $t_s^{\prime(1)}$:  $\tan\phi_v\left(t_s^{\prime(1)}\right) = \frac{v_y\left(t_s^{\prime(1)}\right)}{v_x\left(t_s^{\prime(1)}\right)}$. The ionization rate is proportional to the imaginary part of the tunneling angle $w \propto e^{2m\Im[\phi_v(t_s^{\prime(1)})]}$, where $m$ is the magnetic quantum number. In the case of a spherically symmetric initial state ($s$ state) $m=0$ and the ionization rate does not depend on the tunneling angle, because the electron density in the initial state is the same in all directions. For $p$ states, however, the direction of electron tunneling, defined by the tunneling angle, becomes important. In particular, it leads to the sensitivity of ionization to the sense of rotation of the electron in the initial state. For short-range potentials this effect was predicted and analyzed in \cite{barth2011, barth2013}. In this section we discuss the nonadiabatic Coulomb corrections to the tunneling angle and show how the results in \cite{barth2011, barth2013} are affected by the electron interaction with the long-range core potential.

\begin{figure}
	\includegraphics[scale=0.5]{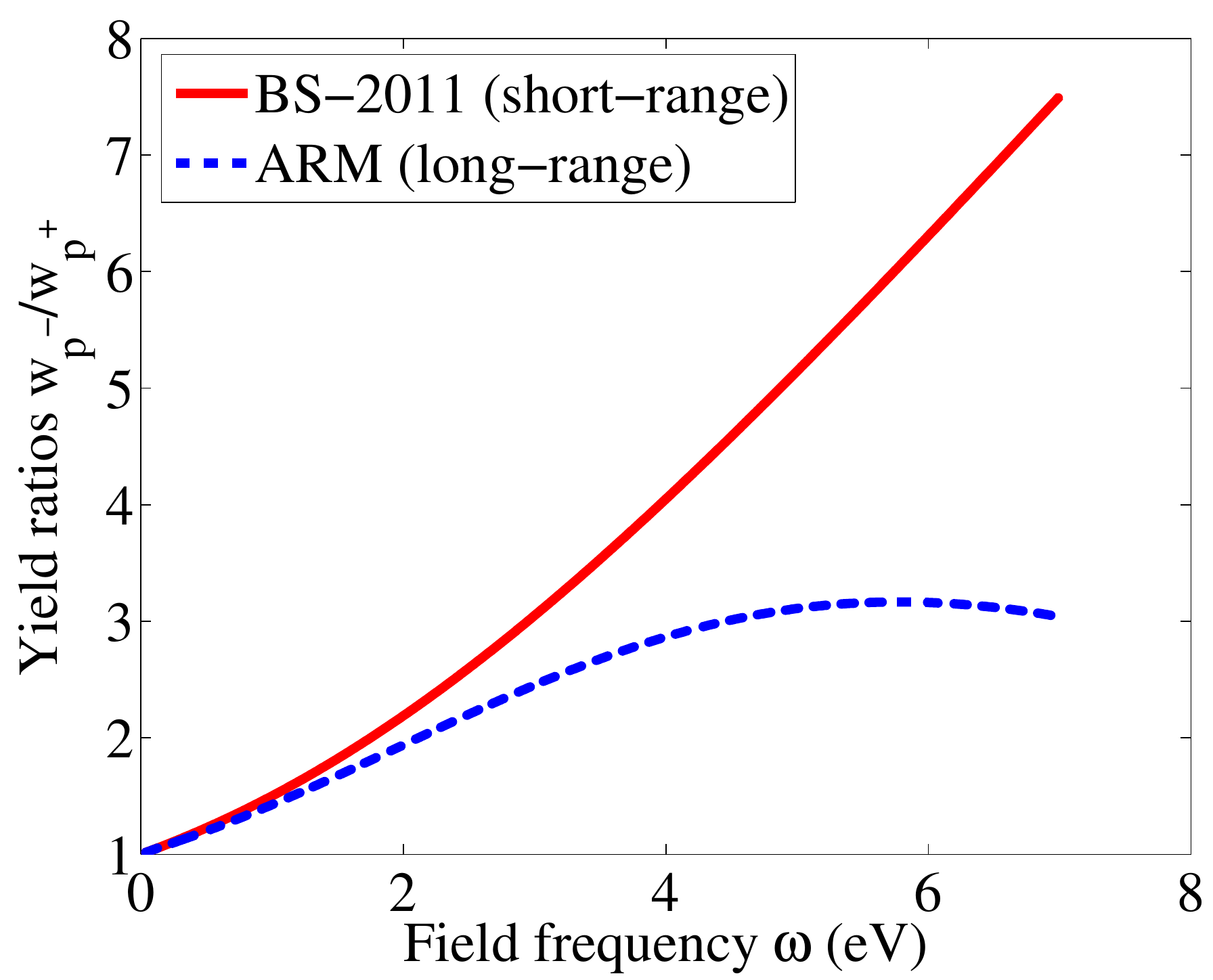}
	\caption{(Color online) Ratio of ionization rates from $p^-$ and $p^+$ orbitals for a Ne atom ($I_p = 21.5645$ eV) and $E_0 = 7.7 \times 10^{10}$ V/m ($E_0 = 0.15$ a.u. and $I = 1.6 \times 10^{15}$ W/cm$^{2}$), with $w_{p^-}/w_{p^+}$ for a right circularly polarized field: short-range potential [solid (red) curve] \cite{barth2011}, and long-range potential [dashed (blue) curve].} \label{fig:ratio}
\end{figure}

The tunneling angle in the case of short-range potentials is
	\begin{equation}
		\tan\phi_v\left(t_s^{\prime(0)}\right) = \frac{p_y - A_0\sin\left(\omega t_s^{\prime(0)}\right)}{p_x - A_0\cos\left(\omega t_s^{\prime(0)}\right)}. \label{phi_v}
	\end{equation}
The Coulomb potential leads to two equally important effects: (i) the modification of the complex ionization time ($t_s^{\prime(0)} + \Delta t_s^{\prime(0)}$ in the long-range potential vs. just $t_s^{\prime(0)}$ in the short-range potential), and (ii) the momentum shift due to the deceleration of the electron by the long-range potential of the core
(see derivation in Sec.~\ref{subsection:boundary_matching}):
	\begin{equation}
		\tan\phi_v^c(t_s')=\frac{v_y\left(t_s^{\prime(0)}\right) - \Delta p_y - \Delta t_s^{\prime(0)} E_y}{v_x\left(t_s^{\prime(0)}\right) - \Delta p_x - \Delta t_s^{\prime(0)} E_x}. \label{tunneling_angle_t_s(1)_again}
	\end{equation}
In this section we focus on the  imaginary part of the complex tunneling angle $\phi_v^c(t_s') = \operatorname{\tan^{-1}}\left(x+iy \right)$, since it contributes to the ionization probability.
The imaginary part of $\phi_v^c(t_s')$ can be cast in the form
	\begin{equation}
		\Im[\phi_v^c(t_s')] = -\frac{1}{4}\ln\left(\left(1-x^2-y^2\right)^2+4x^2\right) + \frac{1}{2}\ln\left(\left(1+y\right)^2+x^2\right). \label{Imtunneling_angle}
	\end{equation}
Note that the real part $x \simeq \mathcal{O}(G_C)$ is of the first order with respect to long-range potential and therefore the $x^2$ terms have to be omitted. The ratio between ionization rates for $p^{-}$ and $p^{+}$ orbitals is
	\begin{gather}
		\frac{w_{p^-}}{w_{p^+}} = \left|\frac{e^{-i2\phi_v^c(t_s^{\prime(1)})}}{e^{i2\phi_v^c(t_s^{\prime(1)})}}\right| = e^{4\Im[\phi_v^c(t_s^{\prime(1)})]} = \left(\frac{1+y}{1-y}\right)^2, \label{ratio}\\
		y = \frac{v_y^{\text{im}} - \Im[\Delta t_s^{\prime(0)}]E_y^{\text{re}}}{v_x^{\text{re}} - \Delta p_x^{\text{re}} + \Im[\Delta t_s^{\prime(0)}]E_x^{\text{im}}}. \label{ratio}
	\end{gather}
Finally,
	\begin{equation}
		y = \frac{v_y^{\text{im}} + {\Delta p_x^{\text{re}}v_x^{\text{re}}}/{\left[p_{\text{opt}}\tanh\omega\tau\right]}}{v_x^{\text{re}}-\Delta p_x^{\text{re}}+{\Delta p_x^{\text{re}}v_x^{\text{re}}}/{p_{\text{opt}}}}. \label{ratio_fin}
	\end{equation}
Figure~\ref{fig:ratio} shows how the nonadiabatic Coulomb effects change the ratio between the ionization rates for the $p^+$ and $p^-$ orbitals. Modifications come solely from the alteration of the tunneling angle. The nonadiabatic Coulomb corrections ($W_{C1}$ and $W_{C2}$) do not contribute to the ratio of the ionization rates, as also noted in \cite{barth2011}. The decrease in the $p^{-}/p^{+}$ ratio at high frequencies in long-range potentials is consistent with the opposite propensity rules in one-photon ionization, where $p^{+}$ is preferred over $p^{-}$ for right circularly polarized fields.

\section{Conclusion} \label{section:conclusion}

We have evaluated strong-field ionization rates and amplitudes for circular fields taking into account nonadiabatic barrier dynamics of a Coulomb potential using the recently developed A$R$M technique. The ionization rates for atoms in arbitrary potentials in circular fields for long-range potentials have been derived rigorously, extending the work in \cite{barth2011} and \cite{lisa2012} and in \cite{ppt1966} and \cite{ppt1967ii}. The A$R$M approach allows for accurate and rigorous analysis of ionization in strong fields, consistently including Coulomb effects both during and after ionization. It should be noted that in the current implementation of the A$R$M method we have included Coulomb effects in first-order perturbation to the action. This limits the applicability of the current implementation to the region of moderate $\gamma$. The simplest ``postmortem" validity check can be performed by computing $\Delta\mathbf{p}^{\text{re}}$ [Eq.~\eqref{Dp_realfin}] and comparing it to the SFA velocities. The momentum shifts $\Delta\mathbf{p}^{\text{re}}$ should not exceed the SFA velocities.


\begin{acknowledgements}
O.S. and J.K. acknowledge support from Marie Curie ITN CORINF. We thank M. Ivanov for many useful discussions, suggestions regarding implementation of nonadiabatic Coulomb effects, and encouragement throughout the work. We thank P. Lambropoulos for many useful discussions including the comments on physics underlying the ionization from the $p^+$ and $p^-$ orbitals. We thank L. Torlina for useful comments.
\end{acknowledgements}

\appendix

\section{Supplementary information for boundary matching} \label{app:boundary}

\subsection{Complex momentum shifts at the boundary} \label{app:momenta}

The goal of this section is to calculate the momentum shift at the matching point $a$,
	\begin{equation}
		\Delta\mathbf{p}(a) = -\int_{t_a^{\prime(0)}}^{T}d\tau\,\nabla U\left(\mathbf{r}_s^{\prime(0)} + \int_{t_a^{\prime(0)}}^{\tau}d\zeta\,\mathbf{v}_{\mathbf{p}}(\zeta)\right), \label{Dp_defA}
	\end{equation}
and show that it does not depend on the position of the boundary under the matching conditions.
We first split the integral into two parts:
	\begin{equation}
		\Delta\mathbf{p}(a) = -\int_{t_a^{\prime(0)}}^{\Re[t_s^{\prime(0)}]}d\tau\,\nabla U\left(\mathbf{r}_s^{\prime(0)} + \int_{t_a^{\prime(0)}}^{\tau}d\zeta\,\mathbf{v}_{\mathbf{p}}(\zeta)\right) - \int_{\Re[t_s^{\prime(0)}]}^{T}d\tau\,\nabla U\left(\mathbf{r}_s^{\prime(0)} + \int_{t_a^{\prime(0)}}^{\tau}d\zeta\,\mathbf{v}_{\mathbf{p}}(\zeta)\right). \label{Dpyreal}
	\end{equation}
Physically, these two parts can be interpreted as accumulated before,
	\begin{equation}
		\Delta\mathbf{p}^{\text{ub}}(a) = -\int_{t_a^{\prime(0)}}^{\Re[t_s^{\prime(0)}]}d\tau\,\nabla U\left(\mathbf{r}_s^{\prime(0)} + \int_{t_a^{\prime(0)}}^{\tau}d\zeta\,\mathbf{v}_{\mathbf{p}}(\zeta)\right),
	\end{equation}
and after,
	\begin{equation}
		\Delta\mathbf{p}^{\text{ic}}(a) = -\int_{\Re[t_s^{\prime(0)}]}^{T}d\tau\,\nabla U\left(\mathbf{r}_s^{\prime(0)} + \int_{t_a^{\prime(0)}}^{\tau}d\zeta\,\mathbf{v}_{\mathbf{p}}(\zeta)\right),
	\end{equation}
where the superscripts ``ub" and ``ic" stand for ``under-the-barrier" and ``in-continuum", respectively. The tunnel exit defined as the coordinate at the time $\Re\left[t_s^{\prime(0)}\right]$,
	\begin{equation}
		\mathbf{r}_e^{\prime(0)} = \int_{t_s^{\prime(0)}}^{\Re[t_s^{\prime(0)}]}d\zeta\,\mathbf{v}_{\mathbf{p}}(\zeta), \label{exitA}
	\end{equation}
is a straightforward extension of Eq.~\eqref{opt_coord_exit}. The second part, $\Delta\mathbf{p}^{\text{ic}}(a)$, does not depend on the boundary. In the following we show that the first part  $\Delta\mathbf{p}^{\text{ub}}(a)$ is negligible under the matching condition $\kappa a \gg 1$.

We first note that $\Delta {p}_{y}^{\text{ub}}(a)$ is purely imaginary, while $\Delta{p}_x^{\text{ub}}(a)$ is purely real. In the same geometry that we use in the text, ${t_a^{\prime(0)}} = i\tau_a^{\prime(0)}$, and the complex under-the-barrier trajectory is $\mathbf{R} = \mathbf{r} + i\boldsymbol{\rho}$:
	\begin{align}
		\mathbf{r} &= -a_0\left[\cosh\phi_i^{\prime(0)} - \cosh\phi\right]\hat{\mathbf{y}} = -a_0\bar{r}\hat{\mathbf{y}},\\
		\boldsymbol{\rho} &= a_0\left[\frac{\phi}{\phi_i^{\prime(0)}}\sinh\phi_i^{\prime(0)} - \sinh\phi \right]\hat{\mathbf{x}} = a_0\bar{\rho}\hat{\mathbf{x}},
	\end{align}
where $\phi_i^{\prime(0)} = \omega\tau_i^{\prime(0)}$, $\phi = \omega\xi$, $a_{0} = A_{0}/\omega$, and $\xi$ is imaginary integration time variable. The Coulomb potential takes the form (details of the analytical continuation of the Coulomb potential to the complex plane will be addressed in our subsequent publication \cite{lisa2013}):
	\begin{equation}
		U\left(\mathbf{R}\right) = -\frac{Q}{\sqrt{r^2-\rho^2}}.
	\end{equation}
The purely imaginary $\Delta p_y^{\text{ub}}(a)$ is
	\begin{equation}
		\Delta p_y^{\text{ub}}(a) = i\frac{Q\omega}{A_0^2}\int_{\phi_a^{\prime(0)}}^{0}\frac{\bar{r}\,d\phi
}{\left(\bar{r}^2 - \bar{\rho}^2\right)^{3/2}}.
	\end{equation}
The purely real $\Delta p_x^{\text{ub}}(a)$ is
	\begin{equation}
		\Delta p_x^{\text{ub}}(a) = -\frac{Q\omega}{A_0^2}\int_{\phi_a^{\prime(0)}}^{0}\frac{\bar{\rho}\,d\phi}{\left(\bar{r}^2 - \bar{\rho}^2\right)^{3/2}},
	\end{equation}
and in both cases, $\phi_a = \omega\tau_a^{\prime(0)}$. Also, since for the optimal trajectory $r \gg \rho$,
	\begin{equation}
		\Delta p_x^{\text{ub}}(a) \simeq -\frac{Q\omega}{A_0^2}\int_{\phi_a^{\prime(0)}}^{0}\frac{\bar{\rho}\,d\phi}{\bar{r}^{3}}.
	\end{equation}
As $\rho = 0$ at the tunnel entrance $\left(\phi = \phi_s^{\prime(0)} = \omega t_s^{\prime(0)}\right)$ and $\rho = 0$ at the tunnel exit ($\phi = 0$), the integral is accumulated in the vicinity of $\tau_a^{\prime(0)}$. We make linear expansion of the integrand around this point,
	\begin{equation}
		\Delta p_x^{\text{ub}}(a) \simeq v_x^{\text{re}}\left(t_s^{\prime(0)}\right)\int_{0}^{\tau_a^{\prime(0)}}d\xi\frac{\tau_a^{\prime(0)}-\xi}{\left\{\kappa\left(\tau_a^{\prime(0)} - \xi\right)+a\right\}^3} = -C\frac{v_x^{\text{re}}\left(t_s^{\prime(0)}\right)}{\kappa}\frac{Q}{\kappa a},
	\end{equation}
where $C$ is a numerical factor:
	\begin{equation}
		C = \int_0^{\infty} \frac{z dz}{(z+1)^3}.
	\end{equation}

So far we have considered $\Delta\mathbf{p}(a)$ defined through its outer-region value. We can also  estimate $\Delta\mathbf{p}(a)$ using its inner-region value. The inner region value of $\Delta\mathbf{p}(a)$ can be calculated using a static approximation (or short-time propagation), since the time interval from $t_s^{\prime(0)}$ to $t_a^{\prime(0)}$ is very small.
It is convenient to estimate $\Delta p_y^{u}(a)$ by evaluating its inner region value.
In a static field, the momentum in the inner region $p^{\text{in}}_y(a)$ is defined through the energy conservation:
	\begin{equation}
		-I_p = \frac{(p^{\text{in}}_y(a))^2}{2} - \frac{Q}{a} - E_0 a.
	\end{equation}
Thus, $p^{\text{in}}_y(a) = -i\sqrt{2(I_p - E_0 a - Q/a)} \simeq - i\sqrt{2(I_p - E_0 a)}(1 + Q/(2a(I_p - E_0a)))$, yielding $p^{\text{in}}_y(a) = -i\kappa(a) - iQ/\kappa(a)a$. The first term is the SFA velocity at the boundary $\kappa(a) = \sqrt{2(I_p-E_0 a)} \simeq \kappa$; the second term is the respective correction associated with Coulomb effects. Thus, $\Delta p_y^{\text{in}}(a) \simeq \mathcal{O}(1/\kappa a)$. The vanishingly small value of the correction at the boundary is not surprising, since the boundary is placed in the region where the Coulomb modification to the barrier is already very small.

\subsection{Additional expressions for boundary matching} \label{subapp:additional}

We derive here the relation:
	\begin{equation}
		j_{\ell}\left(av_{\mathbf{p}^c}\left(t_a^{\prime(1)}\right)\right)e^{-i\mathbf{r}_s^{\prime(0)}\cdot\Delta\mathbf{p}} = j_{\ell}\left(av_{\mathbf{p}}\left(t_s^{\prime(0)}\right)\right). \label{j_l_matchingA}
	\end{equation}
Since the saddle point $t_s^{\prime(1)}$ is close to the SFA saddle point $t_s^{\prime(0)}$, we know that the argument of $j_{\ell}$ is of the order of $\kappa a \gg 1$. So using the large-argument approximation for the spherical Bessel function, and expanding $v_{\mathbf{p}^c}(t')$ up to first order in $\Delta\mathbf{p}$, we get:
	\begin{equation}
		j_{\ell}\left(av_{\mathbf{p}^c}\left(t_a^{\prime(1)}\right)\right) = j_{\ell}\left(av_{\mathbf{p}}\left(t_a^{\prime(1)}\right)\right)e^{a\mathbf{v}_{\mathbf{p}}
\left(t_a^{\prime(1)}\right)\cdot\Delta\mathbf{p}/v_{\mathbf{p}}\left(t_a^{\prime(1)}\right)}.
	\end{equation}
It can be shown that
	\begin{equation}
		j_{\ell}\left(av_{\mathbf{p}}\left(t_a^{\prime(1)}\right)\right) = j_{\ell}\left(av_{\mathbf{p}}\left(t_a^{\prime(0)}\right)\right)\left[1-\frac{a\omega}{\kappa}\Delta t_s^{\prime(0)}\sqrt{(\zeta^2+\gamma^2)(1+\gamma^2)}\right].
	\end{equation}
Since the inner region should be treated in the quasistatic approximation, the second term is vanishingly small.

Analogous to the boundary matching approximation made in \cite{murray2010} we obtain:
	\begin{equation}
		v_{\mathbf{p}}\left(t_a^{\prime(0)}\right)\simeq v_{\mathbf{p}}\left(t_s^{\prime(0)}\right).
	\end{equation}
Taking into account that by definition
	\begin{equation}
		\mathbf{r}_s^{\prime(0)}=a\frac{\mathbf{v}_{\mathbf{p}}\left(t_a^{\prime(0)}\right)}{v_{\mathbf{p}}\left(t_a^{\prime(0)}\right)},
	\end{equation}
and
	\begin{align}
		a\frac{\mathbf{v}_{\mathbf{p}}\left(t_a^{\prime(1)}\right)\cdot\Delta\mathbf{p}}{v_{\mathbf{p}}\left(t_a^{\prime(1)}\right)} &= a\frac{\mathbf{v}_{\mathbf{p}}\left(t_a^{\prime(0)}\right)\cdot\Delta\mathbf{p}}{v_{\mathbf{p}}\left(t_a^{\prime(0)}\right)} + \mathcal{O}(G_C^3),
	\end{align}
we obtain Eq.~\eqref{j_l_matchingA}. It must be noted that it was because of Eq.~\eqref{j_l_matchingA} that $j_{\ell}\left(av_{\mathbf{p}}(t')\right)$ was used in Eq.~\eqref{saddle_t}, and not $j_{\ell}\left(av_{\mathbf{p}^c}(t')\right)$.

\section{Frequency-domain approach} \label{app:freq_domain}

\subsection{The wave function} \label{subapp:wave function}

In our frequency-domain approach we start the analysis with the expression for the wave function in the coordinate representation $\psi(\mathbf{r},t)=\langle \mathbf{r}|\psi_{\text{out}}(t)\rangle$, where $|\psi_{\text{out}}(t)\rangle$ is given by Eq.~\eqref{non-homogeneous}:
	\begin{equation}
		\psi_{\text{out}}(\mathbf{r},t) = i\int_{t_0}^tdt' \int d \mathbf{r}' \int d \mathbf{r}'' \,\langle \mathbf{r}|U_B(t,t')|\mathbf{r}'\rangle \langle \mathbf{r'}|\hat{L}^{-}(a)|\mathbf{r}''\rangle\langle \mathbf{r}''|\psi_{\text{in}}(t')\rangle. \label{psi_t}
	\end{equation}
Taking into account the explicit form of the Bloch operator in coordinate representation Eq.~\eqref{L_coord} we can rewrite Eq.~\eqref{psi_t} as follows:
	\begin{equation}
		\psi(\mathbf{r},t) = i\int_{t_0}^tdt' \int d \mathbf{r}' G_B(\mathbf{r},t;\mathbf{r'},t')\delta(r'-a)B(a,\theta',\phi',t'). \label{psi_tdef1}
	\end{equation}
After following the arguments in Sec.~\ref{subsection:boundary}, we can approximate the boundary term $B(a,\theta',\phi',t')$ as in Eq.~\eqref{boundary}. This follows from the foresight that when we use the saddle point for the time integral, we end up with studying the dynamics of the wave function around the pole $v(t_s') = i\kappa$ in the momentum space. This corresponds to a prominent contribution only from the asymptotic part of the wave function in the region $\kappa r \gg 1$ in coordinate space.

Using Eq.~\eqref{boundary} and Eq.~\eqref{quadratic_expansion} and evaluating the Delta function over $r'$, we now have for the wave function $\psi(\mathbf{r},t)$
	\begin{equation}
		\begin{split}
			\psi(\mathbf{r},t) &= \frac{i\kappa a^2}{(2\pi)^3}\int_{t_0}^tdt'\int_{}^{}d\mathbf{k}\int_{0}^{\pi}d\theta'\int_{0}^{2\pi}
d\phi'\,e^{i(\mathbf{v}_{\mathbf{k}}(t)\cdot\mathbf{r} - \mathbf{v}_{\mathbf{k}}(t')\cdot\mathbf{a})}e^{-\frac{i}{2}\int_{t'}^{t}d\tau\,v^2(\tau)}\varphi_{\kappa\ell}(a)\\
&e^{-i\int_{T}^{t}d\tau\,U(\mathbf{r}_L(\tau;\mathbf{r},\mathbf{k},t)) - iG_C\left(\mathbf{p},T; \mathbf{r}_s^{\prime(0)},t'\right) - \left(\mathbf{a} - \mathbf{r}_s^{\prime(0)}\right)\cdot\nabla G_C\left(\mathbf{p},T; \mathbf{r}_s^{\prime(0)},t'\right)}N_{\ell m}P_{\ell}^m(\cos\theta')e^{im\phi'}, \label{phi'-integral}
		\end{split}
	\end{equation}
where $N_{\ell m} = \sqrt{\frac{2\ell+1}{4\pi}\frac{(\ell-|m|)!}{(\ell+|m|)!}}$ and $\mathbf{a} = a(\sin\theta'\cos\phi'\,\hat{\mathbf{x}} + \sin\theta'\sin\phi'\,\hat{\mathbf{y}} + \cos\theta'\,\hat{\mathbf{z}})$. We also take $\mathbf{A}(t) = -A_0(\cos\omega t\,\hat{\mathbf{x}} + \sin\omega t\,\hat{\mathbf{y}})$.

The point $\mathbf{r}_s^{\prime(0)}$ is defined in spherical coordinates as $\left(a,\theta_s^{\prime(0)},\phi_s^{\prime(0)}\right)$, where $\theta_s^{\prime(0)} = \theta_v(t')$ and $\phi_s^{\prime(0)} = \phi_v(t')$, which gives
	\begin{equation}
		\mathbf{r}_s^{\prime(0)} = a\frac{\mathbf{v}_{\mathbf{k}}(t')}{v_{\mathbf{k}}(t')} \approx \int_{t_s^{\prime(0)}}^{t'}d\tau\,\mathbf{v}_{\mathbf{k}}(\tau). \label{r_s(0)}
	\end{equation}
The approximation follows from the fact that the saddle point for $t'$ will be quite close to the SFA saddle point $t_s^{\prime(0)}$, and hence by defining $t' - t_s^{\prime(0)} = \frac{a}{v_{\mathbf{k}}(t')}$, we can redefine the saddle point $\mathbf{r}_s^{\prime(0)}$ as a classical trajectory. $t'$ corresponds here to a zeroth-order correction in the SFA saddle point when the electron is propagated from a finite boundary instead of the origin [thus the saddle point for $S^{\text{SFA}} + av_{\mathbf{p}}(t')$].

Following Sec.~\ref{surface_integral}, the resulting surface integral is
	\begin{equation}
		I_{\Omega'} = \int_{0}^{\pi}d\theta'\,\sin\theta'\int_{0}^{2\pi}d\phi'\,e^{-i\mathbf{v}(t')\cdot\mathbf{a}}P_{\ell}^m(\cos\theta')e^{im\phi'}e^{i\Delta\mathbf{k}\cdot\mathbf{a}}.
	\end{equation}
The term $e^{i\Delta\mathbf{k}\cdot\mathbf{a}}$ comes from the Taylor expansion of the Coulomb phase $G_C$ about the saddle point coordinate $\left(a,\theta_s^{\prime(0)},\phi_s^{\prime(0)}\right)$. Since the gradient of $G_C$ is identified as the momentum shift, we can see this term as the contribution of the long-range potential to propagation from the finite boundary $r' = a$. Including this shift by rewriting the shifted kinetic momentum as $\mathbf{v}_{\mathbf{k}^c}(t) = \mathbf{v}_{\mathbf{k}} + \mathbf{A}(t) - \Delta\mathbf{k}$, the integral over $\phi'$ is evaluated as
	\begin{equation*}
		\begin{split}
			I_{\phi'}  = \int_{0}^{2\pi}d\phi'\,e^{im\phi'}e^{-i(\mathbf{v}_{\mathbf{k}^c}(t'))\cdot\mathbf{a}} &= \int_{0}^{2\pi}d\phi'\,e^{im\phi'}e^{-av_{\mathbf{k}_{\rho}^c}(t')\sin\theta'\cos(\phi' - \phi_v^c(t')) - ak_z^c\cos\theta'}\\
			& = 2\pi e^{im\phi_v^c(t')}J_m\left(a\mathbf{v}_{\mathbf{k}_{\rho}^c}(t')\sin\theta'\right)e^{-iak_z^c\cos\theta'}.
		\end{split}
	\end{equation*}
The superscript ``$c$" denotes that we are calculating the surface integral over the Coulomb-shifted momentum and $J_n(z)$ is the $n$th-order Bessel function of the first kind.

The $\Omega'$ integral now is:
	\begin{equation}
		\begin{split}
		I_{\Omega'} &= 2\pi(-i)^me^{im\phi_v^c(t')}\int_{0}^{\pi}d\theta'\,J_m\left(av_{\mathbf{k}_{\rho}^c}(t')\sin\theta'\right)P_{\ell}^m\left(\cos\theta'\right)e^{-iak_z^c\cos\theta'}\sin\theta'.
		\end{split}
	\end{equation}
We depart here from the method used in \cite{lisa2012} of approximating the $\theta'$-integral around a given angle according to the direction of polarization (there, $\theta'\sim\pi$ was a reasonable approximation, and here $\theta'\sim\pi/2$). But with the $\theta'\sim\pi/2$ approximation, not only do we lose accuracy in our result, but the small-argument approximation would not be valid for $J_m(ab\sin\theta')$. But we have used $\theta'\sim\pi/2$ for the Coulomb correction, as deviation from a planar trajectory here is suppressed exponentially \cite{popov2004}. Hence we perform an exact analysis, noting that the above integral has an analytic expression from \cite{podolsky1929} by using a similar integral on the product of Bessel functions and the Gegenbauer polynomial from \cite{gnwat1922}, which finally gives us
	\begin{equation}
		I_{\Omega'} = 4\pi(-i)^\ell(-1)^me^{im\phi_v^c(t')}P_{\ell}^m\left(\frac{k_z^c}{v_{\mathbf{k}^c}(t')}\right)j_{\ell}\left(av_{\mathbf{k}^c}(t')\right). \label{theta'}
	\end{equation}
Substituting this result into Eq.~\eqref{theta'} and using Appendix~\ref{subapp:additional}, we get the wave function as:
	\begin{equation}
		\begin{split}
			\psi(\mathbf{r},t) &= N_{lm}(-i)^{\ell}(-1)^m\varphi(a)\frac{2i\kappa a^2}{(2\pi)^2}\int_{t_0}^tdt'\int_{}^{}d\mathbf{k}\,e^{i\mathbf{v}(t)\cdot\mathbf{r}-\frac{i}{2}\int_{t'}^{t}d\tau v^2(\tau)}e^{im\phi_v^c(t')}e^{i\kappa^2(t'-t_0)/2}\\&e^{-i\int_{T}^{t}d\tau\,U(\mathbf{r}_L(\tau;\mathbf{r},\mathbf{k},t))+
iG_C\left(\mathbf{k},T; \mathbf{r}_s^{\prime(0)},t'\right)}P_{\ell}^m\left(\frac{k_z^c}{v_{\mathbf{k}^c}(t')}\right)j_{\ell}\left(av_{\mathbf{k}}(t')\right). \label{wave function}
		\end{split}
	\end{equation}
Equation~\eqref{wave function} is an exact expression from the A$R$M model under the PPT approximation.

\subsection{Ionization Rate}

In order to calculate the ionization rate, we need to know the radial current density, $j_\rho(\mathbf{r},t)$, defined as
	\begin{equation}
		j_\rho(\mathbf{r},t) = \frac{i}{2}\left(\psi(\mathbf{r},t)\frac{\partial\psi^*(\mathbf{r},t)}{\partial\rho} - \psi^*(\mathbf{r},t)\frac{\partial\psi(\mathbf{r},t)}{\partial\rho}\right).
	\end{equation}
Following the procedure of \cite{ppt1966}, but noting the changes due to the presence of the Coulomb phase term, we can get the familiar expression
	\begin{equation}
		w(\mathcal{E},\omega) = 2\pi\sum_{n\ge n_0}^{\infty}\int_{}^{}d\mathbf{k}\,|F_n(\mathbf{k}^c, \omega)|^2\delta\left[\frac{1}{2}\left(k^2 + \kappa^2\left(1+\frac{1}{\gamma^2}\right)\right)-n\omega\right], \label{ionizationrate}
	\end{equation}
with
	\begin{equation}
		\begin{split}
			F_n(\mathbf{k}^c,\omega) &= \frac{\omega}{2\pi}\int_{0}^{2\pi}dt'\,F(\mathbf{k}^c,t')e^{in\omega t'}\\
		&= \frac{2\kappa a^2}{(2\pi)^{3/2}}(-i)^{\ell}(-1)^mN_{\ell m}\varphi_{\kappa\ell}(a)\int_{0}^{2\pi}d(\omega t')\,e^{im\phi_v^c(t')}P_{\ell}^m\left(\frac{k_z^c}{v_{\mathbf{k}^c}(t')}\right)\\&j_{\ell}(av_{\mathbf{k}}(t'))e^{-i\frac{k_\rho\kappa}{\omega\gamma}\sin(\omega t' - \phi_k)+in\omega t'}e^{i\int_{T}^{t'}d\tau\,U\left(\mathbf{r}_L\left(\tau;a,\theta_s^{\prime(0)},\phi_s^{\prime(0)},\mathbf{k},t'\right)\right)}.
		\end{split}
	\end{equation}
The Coulomb phase term is the main difference from the result for the short-range potential.

\subsection{Derivation of $F_n(\mathbf{k},\omega)$} \label{subapp:F_n}

Unlike the result for short-range potentials \cite{ppt1966}, we now have an additional term in the exponential oscillations due to $j_{\ell}(av_{\mathbf{k}}(t'))$, along with the Coulomb corrections. Apart from the modified, Coulomb-shifted momentum that is a new result from this analysis, the Coulomb term in the action also includes motion after ionization, introducing a modification of the result in \cite{barth2011, ppt1966, ppt1967ii} and \cite{ppt1967iii}. As discussed in Sec.~\ref{subsection:boundary_matching}, we know that the saddle point in time would be such that $v\left(t_a^{\prime(0)}\right)\approx\pm i\kappa$, and as done there, we can use the asymptotic condition for a large argument ($\kappa a\gg1$) on the spherical Bessel function:
	\begin{equation}
			j_\ell\left(av_{\mathbf{k}}(t')\right) = \frac{1}{2av_{\mathbf{k}}(t')}\left(e^{i(av_{\mathbf{k}}(t')-(l+1)\pi/2)}+e^{-i(av_{\mathbf{k}}(t')-(l+1)\pi/2)}\right).
	\end{equation}
The two terms correspond to contributions from the diametrically opposite points on the boundary surface $a$, from where we propagate the electron outwards. The point farther from the detector by a distance of $2a$ compared to the point nearer causes an additional exponential decay for propagation from the former. Such a term did not appear in \cite{lisa2012}, as there saddle-point analysis on the $\mathbf{k}$ integral was used, thus isolating the electron field to one particular trajectory, corresponding to a classical particle motion rather than field evolution. Not using the saddle point in our case will naturally lead to interference effects between the contribution from the two points, but under the given condition ($\kappa a\gg1$) those effects will be exponentially small. This way, an interference will be produced on every point throughout every circular disk for different $\theta$ on the sphere $r'=a$. The contribution of each is weighed by the momentum distribution, encoded in $e^{im\phi_v^c(t')}P_l^m\left(\frac{k_z^c}{v_{\mathbf{k}^c}(t')}\right)$. The maximum contribution comes from the region around the saddle point, which effectively considers the electron as a particle. However, since our analysis is exact, the contribution from momenta about the classical are also included in the above result, as well as taking into account the case for nonzero perpendicular momentum ($k_z\neq0$).

The saddle point corresponding to the boundary-dependent action $S_a^{\text{SFA}} = S^{\text{SFA}} + av_{\mathbf{p}}(t')$ can be derived after Taylor expansion about the SFA saddle point $t_s^{\prime(0)}$:
	\begin{equation}
		t_a^{\prime(0)} = t_s^{\prime(0)} - i\frac{a}{\kappa}.
	\end{equation}
After modifying the SFA saddle point $t_s^{\prime(0)}$ through the change in $t_a^{\prime(0)}$ due to the Coulomb phase term, as discussed in Sec.~\ref{section:time_domain}, we get the final expression for the $n$-photon transition amplitude, to first order in $a$:
	\begin{equation}
		\begin{split}
			F_n(\mathbf{k},\omega) &= \frac{a\varphi(a)}{(2\pi)^{3/2}}(-i)^{\ell}(-1)^m\sqrt{\frac{2\pi}{\left|S''\left(t_s^{\prime(0)}\right)\right|}}N_{\ell m}e^{-iS_0\left(t_s^{\prime(0)}\right)}P_\ell^m\left(\frac{k_z^c}{v_{\mathbf{k}^c}\left(t_s^{\prime(1)}\right)}\right)e^{im\phi_v^c\left(t_s^{\prime(1)}\right)}
\\&j_{\ell}\left(av_{\mathbf{k}}\left(t_s^{\prime(1)}\right)\right)
e^{i\int_{T}^{t_a^{\prime(0)}}d\tau\,U\left(\mathbf{r}_L\left(\tau;a,\theta_s^{\prime(0)},\phi_s^{\prime(0)},\mathbf{k},t_s^{\prime(0)}\right)\right)}. \label{F_n(k,w)}
		\end{split}
	\end{equation}
After boundary matching (Sec.~\ref{subsection:boundary_matching}),
	\begin{equation}
		\begin{split}
			F_n(\mathbf{k},\omega) &= \frac{C_{\kappa\ell}N_{\ell m}}{2\pi}(-1)^m\left(1 + (-1)^{\ell+1}e^{-2\kappa a}\right)\sqrt{\frac{\omega\gamma}{k_{\rho}\sqrt{\eta^2-1}}}e^{-iS_0\left(t_s^{\prime(0)}\right) + im\phi_v^c\left(t_s^{\prime(1)}\right)}\\&P_{\ell}^m\left(\frac{k_z^c}{v_{\mathbf{k}^c}\left(t_s^{\prime(1)}\right)}\right)e^{-i\int_{t_{\kappa}^{\prime(0)}}^{T}d\tau\,U\left(\int_{t_s'}^{\tau}d\zeta\,\mathbf{v}(\zeta)\right)}.
		\end{split}
	\end{equation}
Since we are interested in $|F_n(\mathbf{k},\omega)|$ only, we get
	\begin{equation}
		\begin{split}
			\left|F_n(\mathbf{k},\omega)\right|^2 &= \left|C_{\kappa\ell}\right|^2\frac{\omega\gamma}{k_\rho}\frac{2\ell+1}{16\pi^3}\frac{(\ell-|m|)!}{(\ell+|m|)!}\frac{\left(1-(-1)^{\ell}e^{-2\kappa a}\right)^2}{\sqrt{\eta^2-1}}\left|P_{\ell}^m\left(\frac{k_z^c}{v_{\mathbf{k}^c}\left(t_s^{\prime(1)}\right)}\right)\right|^2\\&e^{-2m\Im\left[\phi_v^c\left(t_s^{\prime(1)}\right)\right]}e^{-2\frac{A_0k_{\rho}}{\omega}\left(\eta\operatorname{\cosh^{-1}\eta} - \sqrt{\eta^2-1}\right)}e^{2W_{C1}+2W_{C2}}.
		\end{split}
	\end{equation}
For short-range potentials ($U=0$) the above result matches Eq.~(17) in \cite{barth2011} precisely.

We see another advantage of the A$R$M method here: we now do not have a complicated radial $r'$ integral and the corresponding higher order pole in the momentum-space representation of the wave function. The upshot of the analysis in short-range potentials \cite{barth2011} was that the pole in the momentum-space representation of the wave function was canceled with the zero in the momentum integral at the same point $v(t_s') = i\kappa$. However, for wave functions corresponding to long range potentials, we would have had a $(Q/\kappa+1)$-order pole in the momentum space, leaving a $(Q/\kappa)$-order pole in the final momentum integral. Using the A$R$M method, the Bloch operator isolates the wave function at the boundary $r'=a$ through a $\delta$ function, making that integral straightforward, thus bypassing the pole encountered if the integral was performed over the whole radial domain. At the same time we also get a more robust result, taking into account the Coulomb correction for the ionization rate both during and after ionization.

\subsection{N-Photon ionization Rate} \label{subapp:n_ionization_rate}

The $n$-photon ionization rate is
	\begin{equation}
		\begin{split}
			w_n(\mathcal{E},\omega) &= 2\pi\int_{}^{}d\mathbf{k}\,\left|F_n(\mathbf{k},\omega)\right|^2\delta\left[\frac{1}{2}\left(k^2+\kappa^2\left(1+\frac{1}{\gamma^2}\right)\right)-n\omega\right]\\
				&= \left|C_{\kappa\ell}\right|^2\omega\kappa\frac{2\ell+1}{8\pi^2}\frac{(\ell-|m|)!}{(\ell+|m|)!}\left(1-(-1)^{\ell}e^{-2\kappa a}\right)^2\int_{-\infty}^{\infty}dk_z\int_{0}^{2\pi}d\phi_k\int_{0}^{\infty}dk_{\rho}\\&
e^{-2m\Im\left[\phi_v^c\left(t_s^{\prime(1)}\right)\right]}\left|P_{\ell}^m\left(\frac{k_z^c}{v_{\mathbf{k}}\left(t_s^{\prime(1)}\right)}\right)\right|^2\frac{e^{-\frac{2A_0k_{\rho}\eta}{\omega}\left(\tanh^{-1}\sqrt{1-\frac{1}{\eta^2}}-\sqrt{1-\frac{1}{\eta^2}}\right)}}{A_0\eta\sqrt{1-\frac{1}{\eta^2}}}\\&e^{2W_{C1}+2W_{C2}}\delta\left[\frac{1}{2}\left(k^2 + \kappa^2\left(1+\frac{1}{\gamma^2}\right)\right) - n\omega\right].
		\end{split}
	\end{equation}
Using the Delta function, the integral over $k_\rho$ is easily done by substituting $k_\rho = \sqrt{k_n^2 - k_z^2}$, where $k_n^2 = 2n\omega - \kappa^2\left(1+\frac{1}{\gamma^2}\right)$. We modify the definition of $\zeta = \left(\frac{2n_0}{n} - 1\right)$, used in \cite{ppt1967ii} to include the contribution from the trajectory perpendicular to the plane of polarization to give
	\begin{equation}
		\zeta_{\text{eff}} = \frac{2n_0^{\text{eff}}}{n}-1,
	\end{equation}
where $2n_0^{\text{eff}}\omega = \kappa^2_{\text{eff}}\left(1+\frac{1}{\gamma_{\text{eff}}^2}\right)$, $\kappa_{\text{eff}}^2 = \kappa^2+k_z^2$, and $\gamma_{\text{eff}} = \kappa_{\text{eff}}/A_0$ as defined before. The corresponding values for different functions of $\mathbf{k}$ appearing above are as follows:
	\begin{gather}
		\eta(\mathbf{k}_n) = \sqrt{\frac{1+\gamma_{\text{eff}}^2}{1-\zeta_{\text{eff}}^2}},\\
		\sqrt{1-\frac{1}{\eta^2(\mathbf{k}_n)}} = \sqrt{\frac{\zeta_{\text{eff}}^2+\gamma_{\text{eff}}^2}{1+\gamma_{\text{eff}}^2}},\\
			k_{\rho n} = \sqrt{n\omega (1-\zeta_{\text{eff}})},\\
			A_0 = \sqrt{\frac{n\omega (1+\zeta_{\text{eff}})}{1+\gamma_{\text{eff}}^2}},\\
		\frac{A_0 k_{\rho n}\eta({\mathbf{k}_n})}{\omega} = n = \frac{2n_0^{\text{eff}}}{1+\zeta_{\text{eff}}} = \frac{2n_0}{1+\zeta}.
	\end{gather}
For $k_z \ll k$, we can make the approximation
	\begin{equation}
		\begin{split}
			\operatorname{\tanh^{-1}}\sqrt{1-\frac{1}{\eta^2}} - \sqrt{1-\frac{1}{\eta^2}} &= \frac{1}{2}\ln{\frac{1+\sqrt{1-\frac{1}{\eta^2}}}{1-\sqrt{1-\frac{1}{\eta^2}}}} - \sqrt{1-\frac{1}{\eta^2}}\\
			&\approx\operatorname{\tanh^{-1}}\sqrt{\frac{\zeta^2+\gamma^2}{1+\gamma^2}}-\sqrt{\frac{\zeta^2+\gamma^2}{1+\gamma^2}} + \sqrt{\frac{\zeta^2+\gamma^2}{1+\gamma^2}}\frac{k_z^2}{2k_n^2}.
		\end{split}
	\end{equation}

And since we are comparing our result with \cite{barth2011}, we make the following approximation on the Coulomb-corrected angle $\phi_v^c$: as the corrections $\Delta k_x$ and $\Delta k_y$ are generally small, we can expand to first order in these deviations to write $\phi_v^c$ as a sum of the SFA velocity phase $\phi_v$, and a small correction $\delta$ defined as
	\begin{equation}
		\tan\delta = \frac{\epsilon\tan\phi_v}{1+(1+\epsilon)\tan\phi_v},
	\end{equation}
where $\epsilon = \frac{\Delta k_x}{v_x} - \frac{\Delta k_y}{v_y}$. This way we can split the exponential $e^{-2m\Im\left[\phi_v^c\left(t_s^{\prime(1)}\right)\right]}$,
	\begin{equation}
		e^{-2m\Im\left[\phi_v^c\left(t_s^{\prime(1)}\right)\right]} = e^{-2m\Im\left[\phi_v\left(t_s^{\prime(1)}\right)\right]}
e^{-2m\Im\left[\delta\left(t_s^{\prime(1)}\right)\right]}.
	\end{equation}
A further expansion of $\phi_v\left(t_s^{\prime(1)}\right)$ can be achieved around $\Delta t_s^{\prime(0)}$ to get
	\begin{equation}
		\begin{split}
			e^{-2m\Im\left[\phi_v\left(t_s^{\prime(1)}\right)\right]} &= e^{-2m\Im\left[\phi_v\left(t_s^{\prime(0)}\right)\right]}\exp\left[-2m\Im\left\{\frac{\omega\Delta t_s^{\prime(0)}}{\gamma^2}\left(\frac{\zeta - \gamma^2}{1 + \zeta}\right)\right\}\right]\\
			&= \left(\frac{k_{\rho} - A_0e^{-\operatorname{\cosh^{-1}}\eta}}{k_{\rho} - A_0e^{\operatorname{\cosh^{-1}}\eta}}\right)^m\exp\left[-2m\Im\left\{\frac{\omega\Delta t_s^{\prime(0)}}{\gamma^2}\left(\frac{\zeta_{\text{eff}} - \gamma_{\text{eff}}^2}{1 + \zeta_{\text{eff}}}\right)\right\}\right]. \label{appendix_phi_expansion}
		\end{split}
	\end{equation}

As the probability of escape of the electron in the direction perpendicular to the field is exponentially suppressed, we can make the approximation $k_z\ll k_n$, which gives us
	\begin{equation}
		\begin{split}
			&\left(\frac{k_\rho-A_0e^{-\operatorname{\operatorname{\cosh^{-1}}}\eta}}{k_\rho-A_0e^{\operatorname{\operatorname{\cosh^{-1}}}\eta}}\right)^m \approx\left[\frac{-\zeta - (1-\zeta)\frac{k_z^2}{k_n^2} + \sqrt{\frac{\zeta^2+\gamma^2}{1+\gamma^2}}\left(1+\frac{\varepsilon(k_z)}{2}\right)}{-\zeta -(1-\zeta)\frac{k_z^2}{k_n^2} - \sqrt{\frac{\zeta^2+\gamma^2}{1+\gamma^2}}\left(1+\frac{\varepsilon(k_z)}{2}\right)}\right]^m\\
			& = (-1)^{|m|}\left(1+\frac{1}{\gamma^2}\right)^{|m|}\frac{1}{(1-\zeta^2)^{|m|}}\left(\sqrt{\frac{\zeta^2+\gamma^2}{1+\gamma^2}}-\zeta\operatorname{sgn}(m)\right)^{2|m|},
		\end{split}
	\end{equation}
to first order in $k_z$ and $\varepsilon(k_z) = \frac{k_z^2}{k_n^2}\left(\frac{1-\zeta^2}{\gamma^2+\zeta^2}\right)$.

The second term in Eq.~\eqref{appendix_phi_expansion}, when expanded in powers of $k_z$, has a fourth-order dependence on $k_z$:
	\begin{equation}
		\frac{\zeta_{\text{eff}} - \gamma_{\text{eff}}^2}{1 + \zeta_{\text{eff}}} = \frac{\zeta - \gamma^2}{1+\zeta}\left(1 - \frac{k_z^4}{A_0^2(1+\gamma^2)^2}\right).
	\end{equation}

Finally, we are left with
	\begin{equation}
		\begin{split}
			w_n(\mathcal{E},\omega) &= \left|C_{\kappa\ell}\right|^2\frac{\kappa}{n}\frac{2\ell+1}{4\pi}\frac{(\ell-|m|)!}{(\ell+|m|)!}\left(1-(-1)^{\ell}e^{-2\kappa a}\right)^2\left(\sqrt{\frac{\zeta^2+\gamma^2}{1+\gamma^2}}-\zeta\operatorname{sgn}(m)\right)^{2|m|}\\&\left(1+\frac{1}{\gamma^2}\right)^{|m|}\frac{1}{(1-\zeta^2)^{|m|}}e^{-\frac{4n_0}{1+\zeta}\left(\operatorname{\tanh^{-1}}\sqrt{\frac{\zeta^2+\gamma^2}{1+\gamma^2}}-\sqrt{\frac{\zeta^2+\gamma^2}{1+\gamma^2}}\right)}\sqrt{\frac{1+\gamma^2}{\zeta^2+\gamma^2}}e^{-2m\Im\left[\delta\left(t_s^{\prime(1)}\right)\right]}\\&e^{-2m\frac{\zeta - \gamma^2}{1+\zeta}\Im\left[\frac{\omega\Delta t_s^{\prime(0)}}{\gamma^2}\right]}e^{2W_{C1}+2W_{C2}}\int_{-k_n}^{k_n}dk_z\,e^{-\frac{2n_0}{1+\zeta}\sqrt{\frac{\zeta^2+\gamma^2}{1+\gamma^2}}\frac{k_z^2}{k_n^2}}\left|P_{\ell}^m\left(\frac{k_z}{\pm i\kappa}\right)\right|^2
		\end{split}
	\end{equation}
up to second order in $k_z$. The Coulomb correction is taken out of the integral, on account of its extremely weak dependence on the $k_z$ component of the momentum. The above result is valid for all values of $\ell$ and $m$. An $m$-dependent correction due to the Coulomb potential is also seen to manifest through its effect on the SFA saddle point $t_s^{\prime(0)}$.

To compare with \cite{barth2011}, we consider the case of $\ell=1, m=\pm 1$, for which we have $P_{\ell}^m\left(\frac{k_z}{\pm i\kappa}\right) = -\sqrt{1+\frac{k_z^2}{\kappa^2}}$. To first approximation, we ignore the $\frac{k_z^2}{\kappa^2}$ term in the prefactor, and note that since $n\gg 1$, we can approximate the integral as:
	\begin{equation}
		\int_{-k_n}^{k_n}dk_z\,e^{-n\sqrt{\frac{\zeta^2+\gamma^2}{1+\gamma^2}}\frac{k_z^2}{k_n^2}}\approx \int_{-\infty}^{\infty}dk_z\,e^{-n\sqrt{\frac{\zeta^2+\gamma^2}{1+\gamma^2}}\frac{k_z^2}{k_n^2}} = k_n\sqrt{\frac{\pi}{n}}\left(\frac{1+\gamma^2}{\zeta^2+\gamma^2}\right)^{1/4},
	\end{equation}
which gives
	\begin{equation}
		\begin{split}
			w_n(\mathcal{E},\omega) &= \frac{3\left|C_{\kappa l}\right|^2I_p}{8\sqrt{2\pi} n_0^{3/2}}\frac{e^{2(W_{C1}+W_{C2})}}{\sqrt{1-\zeta}}e^{-\frac{4n_0}{1+\zeta}\left(\operatorname{\tanh^{-1}}\sqrt{\frac{\zeta^2+\gamma^2}{1+\gamma^2}}-\sqrt{\frac{\zeta^2+\gamma^2}{1+\gamma^2}}\right)}e^{-2m\Im\left[\delta\left(t_s^{\prime(1)}\right)\right]}\\&
e^{-2m\frac{\zeta - \gamma^2}{1+\zeta}\Im\left[\frac{\omega\Delta t_s^{\prime(0)}}{\gamma^2}\right]}\left(1+\frac{1}{\gamma^2}\right)^{3/2}\left({\frac{1+\gamma^2}{\zeta^2+\gamma^2}}\right)^{3/4}\left(\sqrt{\frac{\zeta^2+\gamma^2}{1+\gamma^2}}-\zeta\operatorname{sgn}(m)\right)^2. \label{w_n}
		\end{split}
	\end{equation}
The main difference from Eq.~(19) in \cite{barth2011} is the incorporation of Coulomb correction, starting from the tunneling region and into the continuum until the electron is registered at the detector, and an orbital-dependent Coulomb correction, a result that was not expected.

Equation~\eqref{w_n} is equivalent to Eq.~\eqref{ion_rate_opt} obtained within the time-domain approach. However, here we have a result that is valid beyond the optimal momentum, whereas in Eq.~\eqref{ion_rate_opt} we have effectively derived the total ionization rate summed over all photon orders, which is to be compared with Eq.~(6) in \cite{barth2011}. For Eq.~\eqref{w_n}, further discussion of its range requires a knowledge of $\Delta\mathbf{p}$ over all $\mathbf{p}$, and this will be considered elsewhere.

\section{Subcycle ionization amplitude} \label{app:sub_time_domain}

We now consider the case of subcycle ionization amplitudes in time domain, to replace $T \to t$. The subcycle ionization amplitude is defined as
	\begin{equation}
		a_{\mathbf{p}}(t) = i\int_{\mathbf{a}}^{}d\mathbf{r}\,\langle\mathbf{p+A}(t)|\mathbf{r}\rangle\psi(\mathbf{r},t). \label{sub-cycle}
	\end{equation}
Back-propagating the solution $\psi(\mathbf{r},T)$, we can write $\psi(\mathbf{r},t)$ as
	\begin{equation}
		\psi(\mathbf{r},t) = \int_{\mathbf{a}}^{}d\mathbf{r}'\,G(\mathbf{r},t;\mathbf{r}',T)\psi(\mathbf{r}',T) - i\int_{T}^{t}dt'\int_{\mathbf{a}}^{}d\mathbf{r}'\,G(\mathbf{r},t;\mathbf{r}',t')\delta(r'-a)B(a,\theta',\phi',t').
	\end{equation}
The second term represents that part of the wave function that remains bounded within the confines of the Coulomb potential near the atom after ionization. But the wave function content in that region after ionization is negligible compared to the current flux in continuum, thus making the contribution from the former almost 0. So we can write equation \eqref{sub-cycle} as
	\begin{align*}
		a_{\mathbf{p}}(t) &= i\int_{\mathbf{a}}^{}d\mathbf{r}\,\langle\mathbf{p}+\mathbf{A}(t)\vert\mathbf{r}\rangle\int_{\mathbf{a}}^{}d\mathbf{r}'\,G^{\text{EVA}}(\mathbf{r},t;\mathbf{r}',T)\psi(\mathbf{r}',T)\\
	&= i\int_{\mathbf{a}}^{}d\mathbf{r}\int_{\mathbf{a}}^{}d\mathbf{r}'\int_{}^{}d\mathbf{k}\,\frac{e^{-i(\mathbf{p}+\mathbf{A}(t))\cdot\mathbf{r}}}{(2\pi)^{3/2}}\frac{e^{i(\mathbf{k+A}(t))\cdot\mathbf{r}-i\mathbf{k}\cdot\mathbf{r}'}}{(2\pi)^3}e^{-i\int_{T}^{t}d\tau\,U(\mathbf{r}_L(\tau;\mathbf{r},\mathbf{k},t))}\\&e^{-\frac{i}{2}\int_{T}^{t}d\tau\,v^2(\tau)}\psi(\mathbf{r}',T)\\
	&= \frac{1}{(2\pi)^3}\int_{\mathbf{a}}^{}d\mathbf{r}\int_{}^{}d\mathbf{k}\,e^{i(\mathbf{k}-
\mathbf{p})\cdot\mathbf{r}-\frac{i}{2}\int_{T}^{t}d\tau\,v_{\mathbf{k}}^2(\tau)}e^{-i\int_{T}^{t}d\tau\,U(\mathbf{r}_L(\tau;\mathbf{r},\mathbf{k},t))}a_{\mathbf{k}}(T). \label{a_pt_inter}
	\end{align*}
Before we can perform the integration on $\mathbf{r}$, we need to address the $(\mathbf{r},\mathbf{k})$ dependence of the Coulomb correction in the above equation. Similarly to Sec.~\ref{section:time_domain}, we expand the Coulomb phase term $G_C(\mathbf{r},t; \mathbf{k},T) = \int_{T}^{t}d\tau\,U\left(\mathbf{r} + \int_{t}^{\tau}d\zeta\,\mathbf{v}_{\mathbf{k}}(\zeta)\right)$, about the appropriate saddle point $\mathbf{r}_s$ up to quadratic terms in deviation $(\mathbf{a} - \mathbf{r}_s)$. We need the saddle point for the phase term:
	\begin{equation}
		S^{\text{SFA}}(\mathbf{r},\mathbf{k},t) = (\mathbf{k}-\mathbf{p})\cdot\mathbf{r} - \frac{1}{2}\int_{t_s'}^{t}d\tau\,v_{\mathbf{k}}^2(\tau).
	\end{equation}
Therefore,
	\begin{equation}
		\nabla_{\mathbf{k}}S^{\text{SFA}} = 0 \Rightarrow \mathbf{k}_s^{(0)} = \frac{\mathbf{r} - \mathbf{r}_0}{t - t_s^{\prime(0)}}
	\end{equation}
and
	\begin{equation}
		\nabla_{\mathbf{r}}S^{\text{SFA}}\left(\mathbf{r},\mathbf{k}_s^{(0)}(\mathbf{r})\right) = 0 \Rightarrow \mathbf{r}_s^{(0)} = \int_{t_s^{\prime(0)}}^{t}d\tau\,[\mathbf{p} + \mathbf{A}(\tau)].
	\end{equation}
So the classical trajectory can be written as
	\begin{equation}
		\mathbf{r}_s^{(0)} = \int_{t_s'}^{t}d\tau\,\mathbf{v}_{\mathbf{p}}(\tau).
	\end{equation}
After expanding the Coulomb phase term $G_C(\mathbf{r},t; \mathbf{k},T)$ about the saddle points $\left(\mathbf{r}_s^{(0)}, \mathbf{k}_s^{(0)}\right)$ as in Sec.~\ref{section:time_domain}, we can write the subcycle transition amplitude as
	\begin{equation}
		a_{\mathbf{p}}(t) = \frac{1}{(2\pi)^3}\int_{}^{}d\mathbf{k}\int_{}^{}d\mathbf{r}\,e^{i(\mathbf{k} - \mathbf{p})\cdot\mathbf{r} - \frac{i}{2}\int_{T}^{t}d\tau\,v_{\mathbf{k}}^2(\tau) - iG_C(\mathbf{r}_s,t; \mathbf{p},T) - i(\mathbf{r}-\mathbf{r}_s)\cdot\nabla G_C(\mathbf{r}_s,t; \mathbf{p},T)}a_{\mathbf{k}}(T). \label{a_pt_inter}
	\end{equation}
Note the argument $\mathbf{p}$ in $G_C$: the phase term is evaluated for the asymptotic momentum $\mathbf{p}$ and hence the corresponding momentum shift from this Taylor expansion $\Delta\mathbf{p} = -\nabla G_C$ is also evaluated for the asymptotic momentum $\mathbf{p}$ and not for the intermediate momentum $\mathbf{k}$ on which we have to perform the integration.

Following our analysis, we first propagate the electron till the detector after ionization, and to find the momentum shifts at any point of time during this motion, we propagate it back through the EVA Green's function and thus have information on sub-cycle momentum shifts also.

We can now write
	\begin{equation}
		\left.\int_{T}^{t}d\tau\,U\left(\mathbf{r}+\int_{t}^{\tau}d\zeta\,\mathbf{v}_{\mathbf{k}}(\zeta)\right)\right|_{\mathbf{r}=\mathbf{r}_s,\mathbf{k}_s=\mathbf{p}} = \int_{T}^{t}d\tau\,U\left(\int_{t_s'}^{\tau}d\zeta\,\mathbf{v}_{\mathbf{p}}(\zeta)\right).
	\end{equation}
And we can combine this with
	\begin{equation}
		\int_{T}^{t_{\kappa}^{\prime(0)}}d\tau\,U\left(\int_{t_s^{\prime(0)}}^{\tau}d\zeta\,\mathbf{v}(\zeta)\right)
	\end{equation}
in $a_{\mathbf{p}}(T)$ Eq.~\eqref{a_pT}, to get
	\begin{equation}
		\int_{T}^{t}d\tau\,U\left(\int_{t_s^{\prime(0)}}^{\tau}d\zeta\,\mathbf{v}(\zeta)\right) + \int_{t_{\kappa}^{\prime(0)}}^{T}d\tau\,U\left(\int_{t_s^{\prime(0)}}^{\tau}d\zeta\,\mathbf{v}(\zeta)\right) = \int_{t_{\kappa}^{\prime(0)}}^{t}d\tau\,U\left(\int_{t_s^{\prime(0)}}^{\tau}d\zeta\,\mathbf{v}(\zeta)\right),
	\end{equation}
which solves the Coulomb correction for $a_{\mathbf{p}}(t)$. The integral on $\mathbf{r}$ in Eq.~\eqref{a_pt_inter} yields $(2\pi)^3\delta(\mathbf{k} - \mathbf{p} - \Delta\mathbf{p}(t,T))$, and the integral on $\mathbf{k}$ then gives $\mathbf{k} = \mathbf{p} + \Delta\mathbf{p}(t,T)$. The Coulomb  shift $\Delta\mathbf{p}(t,T)$ is now {\it added} instead of being subtracted, which is due to the back-propagation of the electron from the detector with observable $(\mathbf{k},T)$ to $(\mathbf{r},t)$.  We finally get
	\begin{equation}
		\begin{split}
			a_{\mathbf{p}}(t) &= (-1)^{m+1}C_{\kappa\ell}N_{\ell m}\sqrt{\frac{\gamma}{\omega p_{\rho}\sqrt{\eta^2-1}}}e^{-i\int_{t_{\kappa}^{\prime(0)}}^{t}d\tau\,U\left(\int_{t_s'}^{\tau}d\zeta\,\mathbf{v}_{\mathbf{p}}(\zeta)\right)}e^{-\frac{i}{2}\int_{t_s^{\prime(0)}}^{t}d\tau\,v_{\mathbf{p}+\Delta\mathbf{p}}^2(\tau)}
\\&e^{i\kappa^2t_s^{\prime(0)}/2}e^{i\mathbf{r}_s\cdot\Delta\mathbf{p}}
P_{\ell}^m\left(\frac{p_z^c}{v_{\mathbf{p}^c}\left(t_s^{\prime(1)}\right)}\right)e^{im\phi_v^c\left(t_s^{\prime(1)}\right)},
		\end{split}
	\end{equation}
where we have ignored corrections of the order of $\mathcal{O}(G_C^2)$ and greater, which would arise from the Coulomb phase and the Coulomb-shifted velocity phase $\phi_v^c$ after taking $\mathbf{k} = \mathbf{p} + \Delta\mathbf{p}(t,T)$. Expanding $\int_{t_s^{\prime(0)}}^{t}d\tau\,v_{\mathbf{p}+\Delta\mathbf{p}}^2(\tau)$ up to first order in $\Delta\mathbf{p}$, it will cancel the spurious term $\mathbf{r}_s\cdot\Delta\mathbf{p}$. Also, $\mathbf{p}^c$ is defined as $\mathbf{p}^c = \mathbf{p} - \Delta\mathbf{p}\left(t_{a}^{\prime(0)},t\right) \simeq \mathbf{p} - \Delta\mathbf{p}\left(t_i^{\prime(0)},t\right)$, (from discussions in Sec.~\ref{app:momenta}) and hence is boundary independent. The final expression for the sub-cycle transition amplitude is:
	\begin{equation}
		\begin{split}
			a_{\mathbf{p}}(t) &= (-1)^{m+1}C_{\kappa\ell}N_{\ell m}\sqrt{\frac{\gamma}{\omega p_{\rho}\sqrt{\eta^2-1}}}e^{-i\int_{t_{\kappa}^{\prime(0)}}^{t}d\tau\,U\left(\int_{t_s'}^{\tau}d\zeta\,\mathbf{v}_{\mathbf{p}}(\zeta)\right)}e^{-\frac{i}{2}\int_{t_s^{\prime(0)}}^{t}d\tau\,v_{\mathbf{p}}^2(\tau)
+i\kappa^2t_s^{\prime(0)}/2}\\&P_{\ell}^m\left(\frac{p_z^c}{v_{\mathbf{p}^c}\left(t_s^{\prime(1)}\right)}\right)e^{im\phi_v^c\left(t_s^{\prime(1)}\right)}.
\label{a_p(t)}
		\end{split}
	\end{equation}


\section{} \label{app:limit_bessel}
We derive here the result
	\begin{equation}
		\lim\limits_{\rho\to\infty}\frac{\rho J_1(\rho d(\mathbf{k}_{\rho},\mathbf{p}_{\rho}))}{d(\mathbf{k}_{\rho},\mathbf{p}_{\rho})} = 2\pi\delta(\mathbf{k}_{\rho} - \mathbf{p}_{\rho}).
	\end{equation}
We start from the integral
	\begin{equation}
		I_{\rho} = \int_{0}^{2\pi}d\phi\int_{0}^{\rho}d\rho'\,\rho'e^{i(\mathbf{k}_{\rho}-\mathbf{p}_{\rho})\cdot\mathbf{\rho'}}.
	\end{equation}
This integral can be written as
	\begin{align*}
		I_{\rho} &= \int_{0}^{2\pi}d\phi\int_{0}^{\rho}d\rho'\,\rho'e^{i(k_{\rho}\rho'\cos(\phi-\phi_k)-p_{\rho}\rho'\cos(\phi-\phi_p))}\\
		&= \int_{0}^{2\pi}d\phi\int_{0}^{\rho}d\rho'\,\rho'\sum_{n_1=-\infty}^{\infty}i^{n_1}J_{n_1}(k_{\rho}\rho')e^{in_1(\phi-\phi_k)}\sum_{n_2=-\infty}^{\infty}(-i)^{n_2}J_{n_2}(p_{\rho}\rho')e^{-in_2(\phi-\phi_k)}\\
		&= 2\pi\int_{0}^{\rho}d\rho'\,\rho'J_0(\rho' d(\mathbf{k}_{\rho},\mathbf{p}_{\rho})).
	\end{align*}
In going from step 2 to step 3, we first perform the integral over $\phi$ and then use the Graf generalization of Neumann summation. The integral over $\rho'$ is simple:
	\begin{equation}
		\int_{0}^{\rho}d\rho'\,\rho'J_0(\mathbf{k}_{\rho},\mathbf{p}_{\rho}) = \frac{\rho J_1(\rho d(\mathbf{k}_{\rho},\mathbf{p}_{\rho}))}{d(\mathbf{k}_{\rho},\mathbf{p}_{\rho})}.
	\end{equation}
Therefore
	\begin{equation}
		I_{\rho} = 2\pi\frac{\rho J_1(\rho d(\mathbf{k}_{\rho},\mathbf{p}_{\rho}))}{d(\mathbf{k}_{\rho},\mathbf{p}_{\rho})}.
	\end{equation}
Now, by definition,
	\begin{align*}
		(2\pi)^2\delta(\mathbf{k}_{\rho}-\mathbf{p}_{\rho}) &= \int_{0}^{2\pi}d\phi\int_{0}^{\infty}d\rho'\,\rho'e^{i(\mathbf{k}_{\rho}-\mathbf{p}_{\rho})\cdot\mathbf{\rho}'}\\
		&= \lim\limits_{\rho\to\infty}\int_{0}^{2\pi}d\phi_k\int_{0}^{\rho}d\rho'\,\rho'
e^{i(\mathbf{k}_{\rho}-\mathbf{p}_{\rho})\cdot\mathbf{\rho}'}\\
		&= \lim\limits_{\rho\to\infty}2\pi\frac{\rho J_1({\rho d(\mathbf{k}_{\rho},\mathbf{p}_{\rho})})}{d(\mathbf{k}_{\rho},\mathbf{p}_{\rho})}.
	\end{align*}
And we get
	\begin{equation}
		\lim\limits_{\rho\to\infty}\frac{\rho J_1(\rho d(\mathbf{k}_{\rho},\mathbf{p}_{\rho}))}{d(\mathbf{k}_{\rho},\mathbf{p}_{\rho})} = 2\pi\delta(\mathbf{k}_{\rho}-\mathbf{p}_{\rho}),
	\end{equation}
which is the desired result.


\section{} \label{app:equivalence}
We establish the relation
	\begin{equation}
		e^{im\phi_k}\left(\frac{k_{\rho}-A_0e^{-i(\phi_k-\omega t)}}{k_{\rho}-A_0e^{i(\phi_k-\omega t)}}\right)^{m/2} = e^{im\phi_v(t)},
	\end{equation}
where $\phi_v(t) = \operatorname{\tan^{-1}}\left(\frac{v_y(t)}{v_x(t)}\right)$.
We can write
	\begin{equation}
		\begin{split}
			\phi_v(t) &= \tan^{-1}\left(\frac{k_{\rho}\sin\phi_k-A_0\sin\omega t}{k_{\rho}\cos\phi_k-A_0\cos\omega t}\right)\\
		&= \tan^{-1}\left[\frac{k_{\rho}e^{i\phi_k}-A_0e^{i\omega t} - \left(k_{\rho}e^{-i\phi_k}-A_0e^{-i\omega t}\right)}{i\left(k_{\rho}e^{i\phi_k}-A_0e^{i\omega t} + k_{\rho}e^{-i\phi_k}-A_0e^{-i\omega t}\right)}\right].
		\end{split}
	\end{equation}
Taking $\Phi = \left(\frac{k_{\rho}e^{i\phi_k}-A_0e^{\omega t}}{k_{\rho}e^{-i\phi_k}-A_0e^{-i\omega t}}\right)$, we get
	\begin{align*}
		\phi_v(t) = \tan^{-1}\left[i\frac{1-\Phi}{1+\Phi}\right] &= i\tanh^{-1}\left(\frac{1-\Phi}{1+\Phi}\right)\\
		&= \frac{i}{2}\ln\left[\frac{1+\frac{1-\Phi}{1+\Phi}}{1-\frac{1-\Phi}{1+\Phi}}\right]\\
		&= -\frac{i}{2}\ln\Phi = -\frac{i}{2}\ln\left(\frac{k_{\rho}e^{i\phi_k}-A_0e^{i\omega t}}{k_{\rho}e^{-i\phi_k}-A_0e^{-i\omega t}}\right).
	\end{align*}
Therefore,
	\begin{equation}
		\begin{split}
			e^{im\phi_v(t)} &= \exp\left[{\frac{m}{2}\ln\left(\frac{k_{\rho}e^{i\phi_k}-A_0e^{i\omega t}}{k_{\rho}e^{-i\phi_k}-A_0e^{-i\omega t}}\right)}\right]\\
		&=\left(\frac{k_{\rho}e^{i\phi_k}-A_0e^{i\omega t}}{k_{\rho}e^{-i\phi_k}-A_0e^{-i\omega t}}\right)^{m/2} = e^{im\phi_k}\left(\frac{k_{\rho}-A_0e^{-i(\phi_k-\omega t)}}{k_{\rho}-A_0e^{i(\phi_k-\omega t)}}\right)^{m/2},
		\end{split}
	\end{equation}
which is the required result.


%

\end{document}